\documentclass[preprint]{ptephy_v1}

\preprintnumber{arXiv:2307.14018}

\usepackage{graphicx}
\usepackage{dcolumn}
\usepackage{bm}
\usepackage{color}
\usepackage{empheq}

\usepackage{hyperref}
\hypersetup{
colorlinks=false,
hidelinks
}

\let\p\partial
\let\B\bm
\let\f\frac
\def\be{\begin{equation}}\def\ee{\end{equation}}
\def\bea{\begin{eqnarray}}\def\eea{\end{eqnarray}}

\graphicspath{{Figures/}}

\begin{document}

\title{Collective Excitations of Self-Gravitating Bose-Einstein Condensates: Breathing Mode and Appearance of Anisotropy under Self-Gravity}

\author{Kenta Asakawa}
\affil{Department of Physics, Osaka Metropolitan University, 3-3-138 Sugimoto, Sumiyoshi-Ku, Osaka 558-8585, Japan \email{sn22896p@st.omu.ac.jp}}

\author[1,2]{Hideki Ishihara}
\affil{Nambu Yoichiro Institute of Theoretical and Experimental Physics (NITEP), Osaka Metropolitan University, 3-3-138 Sugimoto, Sumiyoshi-Ku, Osaka 558-8585, Japan}

\author[1,2]{Makoto Tsubota}


\begin{abstract}
We investigate the collective mode of a self-gravitating Bose-Einstein condensate (BEC) described by the Gross-Pitaevskii-Poisson (GPP) equations.
The self-gravitating BEC has garnered considerable attention in cosmology and astrophysics, being proposed as a plausible candidate for dark matter.
Our inquiry delves into the breathing and anisotropic collective modes by numerically solving the GPP equations and using the variational method.
The breathing mode demonstrates a reduction in period with increasing total mass due to the density dependence of the self-gravitating BEC, attributed to the density-dependent nature of self-gravitating BECs, aligning quantitatively with our analytical findings.
Additionally, we investigate an anisotropic collective mode in which the quadrupole mode intertwines with the breathing mode. 
The period of the quadrupole mode exhibits similar total mass dependence to that of the breathing mode.
The characteristics of these periods differ from those of a conventional BEC confined by an external potential. 
Despite the differences in density dependence, the ratio of their periods equals that of the BEC confined by an isotropic harmonic potential.
Furthermore, an extension of the variational method to a spheroidal configuration enables the isolation of solely the quadrupole mode from the anisotropic collective mode.
\end{abstract}

\subjectindex{xxxx, xxx}

\maketitle

\section{INTRODUCTION}

Dark matter (DM) stands as a pivotal topic in the realms of cosmology, astrophysics, and modern physics.
The nature and identity of DM have captivated our attention for some decades.
The characteristics of DM entail its non-emissive, non-reflective, and non-absorptive nature towards light while exerting gravitational influence on matter.
Although we cannot directly observe DM in any optical way, evidence of its existence can be inferred from the rotational curve of galaxies \cite{Rubin1970} and gravitational lensing \cite{refregier2003}  \cite{Freese2009}.

The $\Lambda$ cold dark matter ($\Lambda$CDM) model, based on the assumption of non-relativistic DM particles, represents a canonical cosmological framework.
It elucidates the large-scale structures of the universe and has been used for various cosmological and astrophysical studies \cite{Binney}.
However, it is widely acknowledged that the $\Lambda$CDM model has substantial discrepancies at small scales between observational and theoretical results \cite{popolo2017, Bullock2017}.
To address this issue, alternate models have been proposed \cite{banerjee2022}, assuming that, for example, ultralight scalar particles behave as DM \cite{Hui2021, Hu2000} or non-gravitational self-interaction among DM particles exists \cite{spergel2000, Tulin2018}.

An intriguing proposition currently under discussion is the possibility that a DM halo composed of ultralight scalar particles may give rise to a phenomenon known as Bose-Einstein condensation \cite{magana2012,niemeyer2020,ferreira2021}.
Bose-Einstein condensation is a phenomenon in which a macroscopic quantity of Bose particles in a system condenses into a common quantum state.
As a result, the physical state of the Bose-Einstein condensate (BEC) is described by the macroscopic wavefunction.
This phenomenon occurs when the temperature of the Bose gas falls below the critical threshold necessary for Bose-Einstein condensation, indicating that the de Broglie wavelength of Bose particles exceeds the average interparticle spacing \cite{PethickSmith}.

The challenges inherent in the $\Lambda$CDM model could be alleviated through the occurrence of Bose-Einstein condensation.
When the mass of a scalar particle ranges from $m\sim10^{-23}-10^{-22}~\mathrm{eV}$, it is capable of reproducing large-scale structures similar to the $\Lambda$CDM model \cite{ACE,matos2001}.
Furthermore, the Bose gas shows a high critical temperature for Bose-Einstein condensation, typically on the order of $10^{9}~\mathrm{K}$ \cite{silverman2002}, owing to the galactic-scale de Broglie wavelength of approximately $1~\mathrm{kpc}$ \cite{hui2017,niemeyer2020}.
Consequently, the DM halo comprising such ultralight scalar particles has the potential to become a BEC in the Universe.
Due to their ultralight mass, scalar particles manifest their quantum nature on a macroscopic scale within this BEC DM halo, and the uncertainty principle serves to suppress the overpopulation of DM at small scales.
A promising candidate for such ultralight DM is, for example, an axion-like particle proposed in the string theory \cite{marsh2016}.

A BEC of DM is intriguing within the realm of low temperature physics, as it posits the existence of a galaxy-scale quantum fluid beyond the limitations of laboratory settings.
Quantum fluids such as superfluids $^4$He, $^3$He, and atomic BECs are defined as fluids wherein quantum effects macroscopically manifest owing to extremely low temperatures.
Research on these fluids, including collective modes \cite{dalfovo1999}, quantized vortices \cite{fetter2009}, and quantum turbulence \cite{tsatsos2016,tsubota2013}, has been variously conducted in the field of low temperature physics \cite{leggett, PethickSmith, stringari}.
Particularly, a BEC near zero temperature is quantitatively described by the Gross-Pitaevskii (GP) equation, which governs the time evolution of a macroscopic wavefunction \cite{Gross1963, Pitaevskii1961}.
Considering a BEC of ultralight DM, it should be bound by the gravitational potential of the BEC itself, and such a self-gravitating BEC follows the Gross-Pitaevskii-Poisson (GPP) equations \cite{ACE,chavanis2011_1,chavanis2011_2}.

The GPP equations enable DM particles to incorporate a short-range contact interaction.
As mentioned above, such self-interaction of DM particles is proposed to resolve the discrepancies of the $\Lambda$CDM model.
Observational data from colliding clusters of galaxies \cite{harvey2015,jauzac2016} suggest the possibility that DM exhibits non-negligible non-gravitational self-interaction.

A self-gravitating BEC inherently differs from an atomic BEC due to its trapping potential.
In the realm of low temperature physics, a BEC is generally constrained by external potentials, which are predetermined regardless of the configuration or movement of the BEC.
Conversely, a self-gravitating BEC is confined by its gravitational potential, even without any external potential.
Then, the deformation of the self-gravitating BEC can change the depth, range, and anisotropy of the gravitational potential.
These differences between a self-gravitating BEC and an atomic BEC are quite significant when considering the former as a candidate for astrophysical objects, as it causes quantum hydrodynamical phenomena under the influence of self-gravity.

One of the phenomena affected by the anisotropy of the trapping potential is the collective mode. 
It is defined as a low-frequency oscillation of a trapped BEC in response to small fluctuations.
Despite being a many-body system, the collective mode can be described by representative variables as a one-body motion.
For example, within a spherical harmonic potential, a BEC can show the collective mode characterized by the radius, known as the monopole (breathing) mode, and another characterized by two semi-axes or their ratio, termed the quadrupole mode.
An important characteristic is the dependency of collective modes on the shape of the trapping potential; in an axisymmetric harmonic potential, the breathing and quadrupole modes are coupled.
Given the alteration of the gravitational potential with the oscillations of the BEC in self-gravitating systems, the collective mode of such a BEC is likely to exhibit distinct behavior from conventional scenarios.

Various properties of DM halos have been theoretically studied using the GPP equations in recent studies in astrophysics.

The Thomas-Fermi (TF) approximation of the GP equation proves valuable in describing the equilibrium state of a BEC and carrying out the analytical studies.
As for a self-gravitating BEC, this approximation was applied for the first time to compare the consequences with observational data on the rotational curves of galaxies \cite{bohmer2007}.
Subsequently, numerous studies employing this approximation have emerged, including the divergence of the central density in the $\Lambda$CDM model called the core-cusp problem \cite{harko2011}, gravitational lensing effects \cite{harko2015}, rotational deformation \cite{zhang2018}, and the effects of quantized vortices \cite{daller2012, kain2010}.

A self-gravitating BEC has also been investigated through the variational method.
In low temperature physics, the calculation of the collective mode of a BEC using this method is renowned and conventional \cite{PethickSmith,stringari}.
Using the variational method in the study of self-gravitating BECs, the relationship between the total mass and radius of such systems \cite{chavanis2011_1}, gravitational collapse \cite{harko2014,chavanis2016}, and the phase transition between dilute and dense phases \cite{chavanis2018} have been investigated.

Numerical investigations into self-gravitating BECs have recently been reported.
The analytical solution of the GPP equations for general scenarios is intricate.
Hence, numerical calculation of the GPP equations is required to study self-gravitating BECs.
Thus far, the stability of the equilibrium state \cite{guzman2013}, the process of stabilization \cite{guzman2006}, the rotational velocity of a test particle within a self-gravitating BEC \cite{guzman2014,guzman2015}, and the collisions among self-gravitating BECs \cite{gonzalez2011} were numerically investigated by leveraging spatial symmetries to mitigate computational costs.
However, the latest numerical studies have used a three-dimensional system without spatial symmetry, focusing on quantized vortices in self-gravitating BECs, such as their stability \cite{nikolaieva2021, dmitriev2021} and collisions between two self-gravitating BECs with quantized vortices \cite{nikolaieva2022}.

In this work, we study the three-dimensional dynamics of the collective modes of a self-gravitating BEC.
We assume a sufficiently large total mass to employ the TF approximation.
We prepare the initial state by multiplying the factor on the initial velocity with the equilibrium solution of the GPP equations and implement time evolution by numerically solving the GPP equations.
First, the breathing mode of the self-gravitating BEC is induced by radially adding the initial velocity.
Our numerical results agree with the analytical results obtained using the variational method.
Next, by applying the initial velocity axisymmetrically, we numerically obtain the anisotropic collective mode, in which the quadrupole mode is coupled with the breathing mode.

This paper is organized as follows.
Our model and numerical setup are described in Sec.$\mathrm{I}\hspace{-1.2pt}\mathrm{I}$,
Based on this, the breathing mode of a self-gravitating BEC is studied in Sec.$\mathrm{I}\hspace{-1.2pt}\mathrm{I}\hspace{-1.2pt}\mathrm{I}$.
Anisotropic collective mode under self-gravity of the BEC is investigated in Sec.$\mathrm{I}\hspace{-1.2pt}\mathrm{V}$.
Finally, we conclude this paper in Sec.$\mathrm{V}$.

\section{MODEL AND NUMERICAL SETUP}
\subsection{GROSS-PITAEVSKII-POISSON MODEL AND THE EQUILIBRIUM STATE}
We consider a self-gravitating BEC composed of scalar bosons with mass $m$ and an s-wave scattering length $a$ at zero temperature.
In the GPP model, the physical state of the BEC is described by the macroscopic wavefunction $\psi(\B{r},t)$, and the time evolution is given by the following GPP equations defined as
\be
\begin{cases}

i\hbar\p_t{\psi} = -\f{\hbar^2}{2m}\nabla^2\psi+[mV+\f{4\pi\hbar^2a}{m} |\psi|^2]{\psi},

\\

{\nabla}^2V= 4\pi mG\lvert{\psi}\rvert^2,
\end{cases}
\label{gpp}
\ee
where $V(\B{r},t)$ denotes the gravitational potential \cite{ferreira2021,chavanis2011_1,ACE}.
Using the Madelung representation written by
\be
\psi(\B{r},t)=\sqrt{\f{\rho(\B{r},t)}{m}}e^{i\theta(\B{r},t)},
\label{madelung}
\ee
the GP equation yields the continuity equation and the Euler-like equation, where $\rho=m|\psi|^2$ denotes the density and $\B{v}=\hbar\nabla\theta/m$ denotes the velocity field \cite{PethickSmith}.
Then, the total mass is
\be
M=\int d\B{r}\rho(\B{r})=m\int d\B{r}|\psi|^2.
\label{total_mass}
\ee
The total energy $E$ is the sum of the kinetic energy $K$, the potential energy $W$, and the self-interaction energy $I$ given by \cite{kain2010}
\begin{align}
\label{total_energy} 
E&=K+W+I, \\
\label{kinetic_energy}
K&=\f{\hbar^2}{2m}\int d\B{r}|\nabla\psi|^2, \\
\label{potential_energy}
W&=\f{m}{2}\int d\B{r}|\psi|^2V, \\
\label{self-interaction_energy}
I&=\f{2\pi\hbar^2a}{m}\int d\B{r}|\psi|^4.
\end{align}

We can obtain the equilibrium solution of the GPP equations (\ref{gpp}) using the TF approximation \cite{bohmer2007}.
When the kinetic energy is negligible, denoted as $K\ll I$, the configuration of the equilibrium state is formed by the competition between gravity and self-interaction \cite{PethickSmith}.
The kinetic energy becomes insignificant, when the total mass is sufficiently large \cite{harko2019}.
The equilibrium configuration is expected to exhibit spherical symmetry under self-gravity.
Assuming that the equilibrium state satisfies the ansatz $\psi(\B{r},t)=\phi(r)e^{-i\mu t/\hbar}$, where $\mu$ denotes the chemical potential, the steady solution of Eq. (\ref{gpp}) is approximately given by
\be
	\rho(r)
	\simeq
	\rho_cj_0\left(\f{\pi r}{R_\mathrm{TF}}\right) \\
	=
	\rho_c\f{\sin\left(\pi r/R_\mathrm{TF}\right)}{\left(\pi r/R_\mathrm{TF}\right)}~~(r<R_\mathrm{TF}),
\label{tf}
\ee
where $r=|\B{r}|$ represents the radial coordinate, $\rho_c$ stands for the central density, and $j_0$ is the 0-th spherical Bessel function.
Subsequently, the TF radius $R_\mathrm{TF}$ is defined as the minimum radius at which the density becomes zero:
\be
R_\mathrm{TF}=\pi\sqrt{\f{\hbar^2a}{Gm^3}},
\label{tf_radius}
\ee
indicating that the size of a massive self-gravitating BEC remains independent of its total mass.
There are no densities in the range of $r>R_\mathrm{TF}$.
Thus, the cumulative mass profile is expressed as
\be
M\left(r\right)=\begin{cases}
			\f{M_\mathrm{TF}}{\pi}\left\{\sin\left(\f{\pi r}{R_\mathrm{TF}}\right)-\f{\pi r}{R_\mathrm{TF}}\cos\left(\f{\pi r}{R_\mathrm{TF}}\right)\right\} & (r<R_\mathrm{TF}), \\
			M_\mathrm{TF} & (r>R_\mathrm{TF}),
		      \end{cases}
\label{cum_mass}
\ee
where $M_\mathrm{TF}$ is defined as
\be
M_\mathrm{TF}=\f{4\rho_cR_\mathrm{TF}^3}{\pi}.
\label{tf_mass}
\ee
Given that the TF radius remains unaffected by the total mass, Eq. (\ref{tf_mass}) reveals that only the central density escalates, even with an increase in the total mass.
By solving the Poisson equation, the gravitational potential becomes
\be
	V(r)
	=G\int^r_\infty ds\f{M(s)}{s^2} \\
	=\begin{cases}
		-\f{GM_\mathrm{TF}}{R_\mathrm{TF}}\left\{1+j_0\left(\f{\pi r}{R_\mathrm{TF}}\right)\right\} & (r<R_\mathrm{TF}),\\
		-\f{GM_\mathrm{TF}}{r} & (r>R_\mathrm{TF}).
	\end{cases}
\label{g_tf}
\ee
Therefore, using Eqs. (\ref{kinetic_energy}), (\ref{potential_energy}), (\ref{self-interaction_energy}), (\ref{tf}), and (\ref{g_tf}), the energy of each equilibrium state can be expressed as
\begin{align}
\label{kinetic_eq}
&K\simeq\f{\pi}{8}\f{\hbar^2M_\mathrm{TF}}{m^2R_\mathrm{TF}^2}\left\{\mathrm{Si}(\pi)-\pi+\int_0^{\Lambda\pi}dx\f{x}{\sin x}\right\}, \\
\label{potential_eq}
&W=-\f{3}{4}\f{GM_\mathrm{TF}^2}{R_\mathrm{TF}}, \\
\label{self-interaction_eq}
&I=\f{\pi^2}{4}\f{\hbar^2aM_\mathrm{TF}^2}{m^3R_\mathrm{TF}^3},
\end{align}
where $\mathrm{Si}(x)$ is a sine integral and $\Lambda\pi(<\pi)$ serves as the cutoff to avoid the divergence of the integral for $\Lambda\rightarrow 1$.
The divergence is derived from the breakdown of the TF approximation, and Eq. (\ref{kinetic_eq}) shows the kinetic energy within a spherical region whose radius is obtained by subtracting the depth of the BEC surface from the TF radius.
The value of $\Lambda$ can be estimated using the ratio between the coherence length $\xi=\sqrt{m/(8\pi a\rho_c)}$ and the TF radius, such that $\Lambda\sim 1-\xi/R_\mathrm{TF}$.

\subsection{NUMERICAL SETUP}
The GPP equations (\ref{gpp}) are expressed in the following dimensionless formulation: 
\be
\begin{cases}
i\tilde{\p}_t\tilde{\psi}=-\f{1}{2}\tilde{\nabla}^2\tilde{\psi}+[\tilde{V}-i\tilde{V}_\mathrm{s}+\tilde{a}|\tilde{\psi}|^2]\tilde{\psi}, \\
\tilde{\nabla}^2\tilde{V}=|\tilde{\psi}|^2,
\end{cases}
\label{dless_gpp}
\ee
where the tilde symbols represent the dimensionless variables, and as explained later, the term $-i\tilde{V}_\mathrm{s}$ in the dimensionless GP equation is artificially added due to a computational reason. 
The dimensionless variables are defined as follows: $\tilde{\B{r}}=\B{r}(\tilde{\lambda} mc/\hbar), \tilde{t}=t(\tilde{\lambda}^2mc^2/\hbar), \tilde{\psi}=\psi\{\sqrt{4\pi G}\hbar/(\sqrt{m}c^2\tilde{\lambda}^2)\}, \tilde{V}=V/(\tilde{\lambda}^2c^2), \tilde{a}=a\{\tilde{\lambda}^2c^2/(mG)\}$ \cite{guzman2014}.
The scaling factor $\tilde{\lambda}$ can adjust the system size using the invariance of the equations for any value\footnote{We introduce $\tilde{\lambda}$ to keep the values of the dimensionless variables within a suitable range for the numerical calculations.}.
Then, the dimensionless total mass and energy are given by $\tilde{M}=M\{4\pi Gm/(\tilde{\lambda}\hbar c)\}$ and $\tilde{E}=E\{8\pi Gm/(\hbar c^3\tilde{\lambda}^3)\}$.

The boundary conditions should be taken into account when dealing with Eq. (\ref{dless_gpp}).
The GP equation is typically solved under periodic boundary conditions.
However, the Poisson equation is inconsistent with the periodic boundary condition owing to the long-range gravitational interaction.
To overcome this inconsistency, we propose a novel method.

The GP equation in Eq. (\ref{dless_gpp}) is computed using the fourth-order Runge-Kutta and pseudo-spectrum methods under periodic boundary conditions.
To solve the Poisson equation, we use the finite difference and Jacobi methods at each time step \cite{Thijssen}.
Using the Jacobi method, the solution of the Poisson equation in Eq. (\ref{dless_gpp}) manifests as the equilibrium solution of the diffusion equation for the function $\tilde{V}'(\tilde{\B{r}},\tilde{t};\tilde{s})$:
\be
\f{d\tilde{V}'}{d\tilde{s}}(\tilde{\B{r}},\tilde{t};\tilde{s})
=
\tilde{\nabla}^2\tilde{V}'(\tilde{\B{r}},\tilde{t};\tilde{s})
-
|\tilde{\psi}(\tilde{\B{r}},\tilde{t})|^2,
\label{diffusion}
\ee
where $\tilde{s}$ denotes a real parameter.
The gravitational potential is obtained from $\tilde{V}(\tilde{\B{r}},\tilde{t})=\lim_{\tilde{s}\rightarrow\infty}\tilde{V'}(\tilde{\B{r}},\tilde{t};\tilde{s})$.
Furthermore, the efficacy of the Jacobi method can be notably enhanced through Nesterov's accelerated gradient method \cite{nesterov1983,su2016,donoghue2015}.
This method is common in the field of machine learning and enables us to rapidly converge the function $\tilde{V'}(\tilde{\B{r}},\tilde{t};\tilde{s})$ toward the gravitational potential $\tilde{V}(\tilde{\B{r}},\tilde{t})$.
The function $\tilde{V'}(\tilde{\B{r}},\tilde{t};\tilde{s})$ satisfies the Dirichlet boundary condition given by
\be
\tilde{V}'(\tilde{\B{r}}=\tilde{\B{R}}_b,\tilde{t};\tilde{s})
=
-\f{\tilde{M}(\tilde{t})}{4\pi|\tilde{\B{R}}_b-\tilde{\B{R}_c}|}
\label{bc_poisson}
\ee
where $\tilde{\B{R}}_b$ denotes the position on the boundary of the numerical domain and $\tilde{\B{R}_c}$ denotes the center of the numerical domain.
We apply a time-dependent total mass $\tilde{M}(\tilde{t})$ because, as mentioned later, we consider the probability of emission of particles.
The boundary condition (\ref{bc_poisson}) is appropriate when the BEC occupies a region smaller than the numerical domain and is located near its center.

In our numerical model, the imaginary potential $-i\tilde{V}_\mathrm{s}$, called "sponge" potential, written by
\be
\tilde{V}_\mathrm{s}(\B{r})
=
\f{\tilde{V}_o}{2}\left\{2+\tanh\left(\f{\tilde{r}-\tilde{r}_c}{\tilde{\delta}}\right)-\tanh\left(\f{\tilde{r}_c}{\tilde{\delta}}\right)\right\}
\label{sponge}
\ee
is introduced into Eq. (\ref{dless_gpp}) \cite{guzman2014}.
Here, $\tilde{V}_o$ denotes the amplitude.
This sponge potential can reduce the number of particles within the range of $\tilde{r}>\tilde{r}_c$, where $\tilde{\delta}$ denotes the step width.
Functioning as a sink, the sponge potential aids in removing particles emitted from the BEC.
As the BEC undergoes motion, some particles acquire large kinetic energy, leading to their escape from the gravitational potential.
However, in our system, they return to the BEC due to periodic boundary conditions on the GP equation.
To avoid the unphysical situation, the sponge potential becomes imperative.

The initial state is prepared by multiplying the initial phase factor $\exp[i\tilde{\phi}(\tilde{\B{r}})]$ to the equilibrium state $\tilde{\psi}_\mathrm{eq}(\tilde{\B{r}})$:
\be
\tilde{\psi}(\tilde{\B{r}},\tilde{t}=0)
=
\tilde{\psi}_\mathrm{eq}(\tilde{\B{r}})e^{i\tilde{\phi}(\tilde{\B{r}})}.
\label{initial_state}
\ee
The BEC is located at the center of the numerical domain.
We obtain the equilibrium state $\tilde{\psi}_\mathrm{eq}$ using the imaginary time propagation method of the GPP equations.
The initial phase is given by
\be
\tilde{\phi}(\tilde{r}_\perp,\tilde{z})=\f{1}{2}\left(\tilde{\alpha} \tilde{r}_\perp^2+\tilde{\beta} \tilde{z}^2\right),
\label{ini_phase}
\ee
which gives rise to the initial velocity field $\tilde{\B{v}}=\tilde{\nabla}\tilde{\phi}=(\tilde{\alpha} \tilde{r}_\perp)\B{e}_\perp+(\tilde{\beta} \tilde{z})\B{e}_z$.
Here, $\tilde{r}_\perp=\sqrt{\tilde{x}^2+\tilde{y}^2}$ and $\tilde{z}$ denote the 3D cylindrical coordinates, and the unit vectors along each direction are $\B{e}_\perp$ and $\B{e}_z$.
We determine the initial velocity by $\tilde{\alpha}$ and $\tilde{\beta}$ (e.g., it becomes isotropic when $\tilde{\alpha}=\tilde{\beta}$).

Finally, the physical and numerical parameters are specified.
We consider the case in which each boson has a mass $m=3\times10^{-24}~\mathrm{eV}$ and an s-wave scattering length $a=5.62\times10^{-98}\lambda_c~\mathrm{m}$, where $\lambda_c$ denotes the Compton wavelength \cite{nikolaieva2021}.
For the numerical analyses, we use a 3D cube of length $\tilde{L}_x=\tilde{L}_y=\tilde{L}_z=40$.
The scaling factor is $\tilde{\lambda}=3\times10^{-3}$, ensuring that the numerical domain vastly exceeds the size of the BEC.
The spatial grid is configured as $\tilde{N}_x=\tilde{N}_y=\tilde{N}_z=128$.
The time resolution of the GP equation is $\Delta \tilde{t}=10^{-3}$, and the resolution of the parameter $\tilde{s}$ is $0.05\times(\tilde{L}_x/\tilde{N}_x)^2\approx4.88\times 10^{-3}$.
The parameters of the sponge potential are designated as $\tilde{V}_o=1000, \tilde{r}_c=0.8\times(\tilde{L}_x/2)=16$ and $\tilde{\delta}=2\times(\tilde{L}_x/\tilde{N}_x)=0.625$.
Then, $\tilde{r}_c$ exceeds the dimensionless TF radius $\tilde{R}_\mathrm{TF}\approx 9.09$, which represents the typical size of a BEC.
This is reasonable because the sponge potential works outside the BEC.
In all our simulations, the change rate of the total mass from its initial value is less than one percent, and particles are hardly emitted from the BEC.

\section{BREATHING MODE}

In this section, we investigate the breathing modes of self-gravitating BEC.
For a conventional BEC confined by an external potential, the initial radial velocity field induces the breathing mode \cite{PethickSmith,stringari,castin1996}.
Likewise, we analytically and numerically investigate the breathing mode of a self-gravitating BEC by setting a spherically symmetric initial phase.
We validate our numerical results through comparison with our analytical predictions.

\subsection{ANALYTICAL CALCULATION BY VARIATIONAL METHOD}

We apply the variational method in the GPP model to reveal the breathing modes of a self-gravitating BEC.
We examine a self-gravitating BEC with a fixed total mass $M$.
Previous studies have used Gaussian \cite{chavanis2011_1} and quadratic functions \cite{harko2014} as trial functions of the variational method.
We adopt the 0-th spherical Bessel function as the trial function because we posit that the self-gravitating BEC possesses a large total mass, allowing us to use the TF approximation.
This approach is the most suitable for such a large total mass.

We prepare a trial function for the variational method.
When the self-gravitating BEC is so large that the TF approximation is available, the macroscopic wavefunction of the equilibrium state is derived as
\be
\psi_\mathrm{eq}(r)=\sqrt{\f{\pi M}{4mR_\mathrm{TF}^3}j_0\left(\f{\pi r}{R_\mathrm{TF}}\right)}
\label{wavefunc_eq}
\ee
using Eqs. (\ref{tf}) and (\ref{tf_mass}).
Assuming that the density profile maintains its form as in Eq. (\ref{wavefunc_eq}) during the motion of the cloud, the trial function can be obtained as
\be
\psi(\B{r},t)
=
\sqrt{
	\f{\pi M}{4mR(t)^3}
	j_0\left(\f{\pi r}{R(t)}\right)
}
\exp\left[i\f{mr^2}{2\hbar}H(t)\right]  
\label{trial_func}
\ee
in the range of $r<R(t)$ \cite{PethickSmith,harko2014,chavanis2016}, and the radius $R(t)$ and $H(t)$ are treated as variables governing the breathing mode.

We derive the Euler-Lagrange equations for $R(t)$ and $H(t)$.
In the GPP model, the Lagrangian can be described by
\be
 L=\f{i\hbar}{2}\int d\B{r}\left(\psi^*\p_t\psi-\psi\p_t\psi^*\right)-E,
 \label{lagrangian_gpp}
\ee
using Eq. (\ref{total_energy}) \cite{harko2014}.
Upon substituting the trial function (\ref{trial_func}), the Lagrangian transforms into
\be
L(R,H)=-\f{1}{2}\left(1-\f{6}{\pi^2}\right)MR^2(\dot{H}+H^2)-U(R).
\label{lagramgian}
\ee
Here, the effective potential $U(R)$ is defined as
\be
	U(R) = \f{\pi}{8}\f{\hbar^2M}{m^2R^2}F(\Lambda)-\f{3}{4}\f{GM^2}{R}+\f{\pi^2}{4}\f{\hbar^2aM^2}{m^3R^3}
		\equiv \f{C_z}{R^2}-\f{C_p}{R}+\f{C_i}{R^3}.
\label{eff_pot}
\ee
Here the coefficients are $C_z=\f{\pi}{8}\f{\hbar^2M}{m^2}F(\Lambda)$, $C_p=\f{3}{4}GM^2$, and $C_i=\f{\pi^2}{4}\f{\hbar^2aM^2}{m^3}$, and the function $F(\Lambda) =\mathrm{Si}(\pi)-\pi+\int_0^{\Lambda\pi}dx\f{x}{\sin x} $.
Consequently, the Euler-Lagrange equation for $H(t)$ is $H(t)=\dot{R}(t)/R(t)$, and the radius $R(t)$ follows the equation\footnote{These Euler-Lagrange equations imply that the motion of the spherical self-gravitating BEC resembles that of the Friedmann model because Eq. (\ref{eom_r}) corresponds to the Friedmann equation. Here, $R$ and $H$ are the scale factor and the Hubble parameter, respectively.} given by
\be
\left(1-\f{6}{\pi^2}\right)M\f{d^2R(t)}{dt^2}=-\f{dU(R)}{dR}.
\label{eom_r}
\ee

The equilibrium solution $R_\mathrm{eq}$ is acquired by setting $dU(R)/dR=0$, yielding
\be
	R_\mathrm{eq}
	=\f{C_z+\sqrt{C_z^2+3C_pC_i}}{C_p} \\
	=\f{\pi^2\hbar^2}{6Gm^2}\f{F(\Lambda)}{M}\left\{1+\sqrt{1+\f{36Gam}{\pi^2\hbar^2}\f{M^2}{F(\Lambda)^2}}\right\}.
\label{mass_radius}
\ee
\begin{figure}
\centering
\includegraphics[width=8cm]{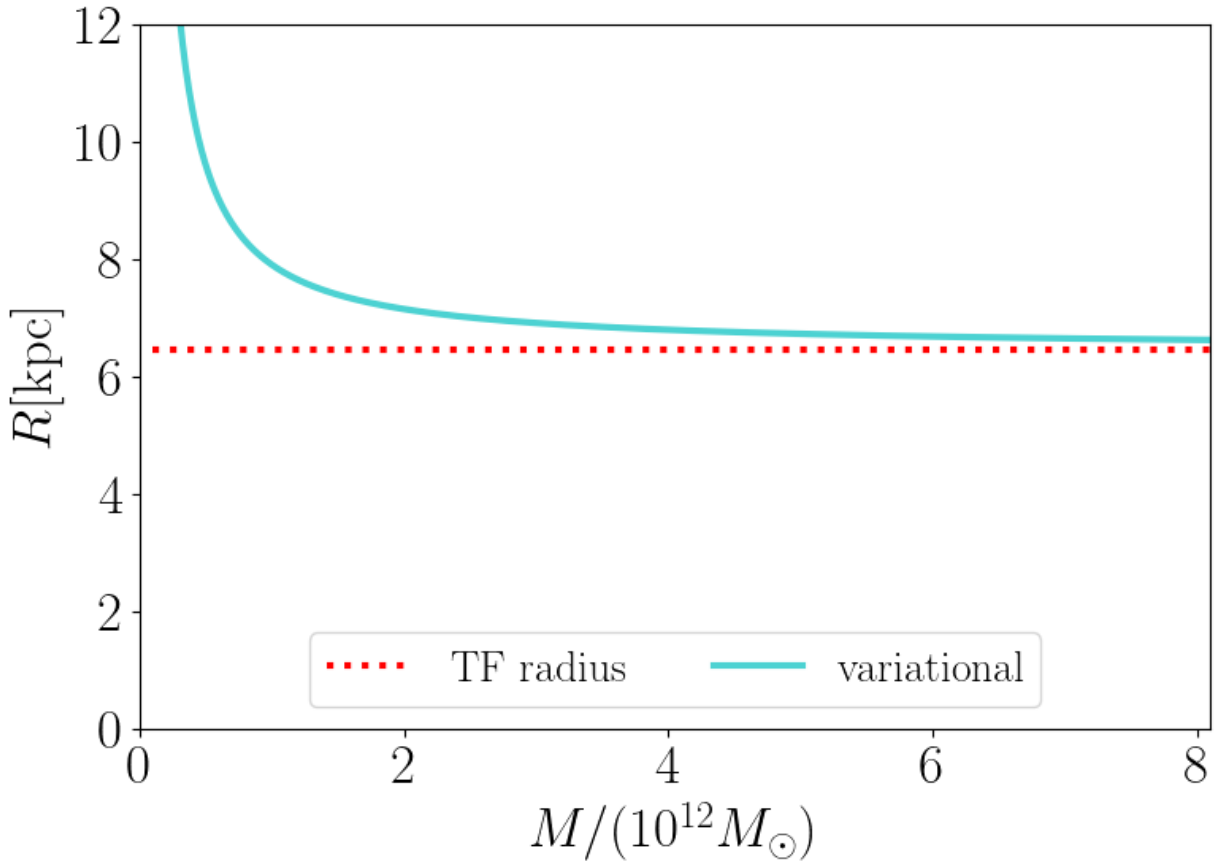}
\caption{
The relationship between the total mass $M$ and the radius $R$ derived from Eq. (\ref{tf_radius}) with the cut-off parameter set to $\Lambda=0.97$.
The horizontal axis shows the total mass, and the vertical axis shows the radius.
The cyan solid line shows the analytical result of Eq. (\ref{tf_radius}), and the red dotted line shows the TF radius $R_\mathrm{TF}$.
}
\label{figure1}
\end{figure}
Fig. \ref{figure1} describes the result of Eq. (\ref{mass_radius}) with $\Lambda=0.97$ as the cut-off parameter.
This figure shows that as the total mass $M$ increases, $R_\mathrm{eq}$ monotonically decreases due to gravity and converges to the TF radius defined in Eq. (\ref{tf_radius}) as $M\rightarrow\infty$.
Thus, our results offer an enhanced understanding of the relationship, particularly for situations with large total masses, compared to previous studies \cite{harko2014,chavanis2011_1} employing other trial functions.

The breathing mode manifests as a small oscillation of radius around $R_\mathrm{eq}$.
The radius is expressed as $R(t)=R_\mathrm{eq}+\delta R(t)$, using its fluctuation $\delta R(t)~(|\delta R(t)|\ll R_\mathrm{eq}$).
Employing Eqs. (\ref{eff_pot}), (\ref{eom_r}), and (\ref{mass_radius}), we obtain the following equation of motion:
\be
	\f{d^2\delta{R}(t)}{dt^2} \simeq -\f{\pi^2}{\pi^2-6}\f{1}{M}\f{d^2U(R_\mathrm{eq})}{dR^2}\delta R(t)
				=-\f{2\pi^2}{\pi^2-6}\f{C_zR_\mathrm{eq}+3C_i}{MR_\mathrm{eq}^5}\delta R(t).
\label{osc_eom_g}
\ee
In this context, we assume the total mass of the BEC to be sufficiently large to warrant the TF approximation.
Given the negligible contribution of $C_z$, Eq. (\ref{osc_eom_g}) is simplified to
\be
\f{d^2\delta{R}(t)}{dt^2}=-\f{3}{2\pi(\pi^2-6)}\sqrt{\f{G^5m^9}{\hbar^6a^3}}M\delta R(t), 
\label{osc_eom}
\ee
which implies a harmonic oscillation with a period
\be
T_\mathrm{B}=2\pi\sqrt{\f{2\pi(\pi^2-6)}{3}\sqrt{\f{\hbar^6a^3}{G^5m^9}}}\f{1}{\sqrt{M}}.
\label{periodicity_breathing}
\ee
The dependence on $M$ has also been delineated in previous studies \cite{harko2014,chavanis2011_1}.
The prefactor and $M$-dependence can be confirmed by the following numerical calculations.

\subsection{NUMERICAL RESULTS}

To confirm Eq. (\ref{periodicity_breathing}), we conduct numerical simulations for the total mass $M/(10^{14}M_\odot)=1, 2, 3, 4$, where $M_\odot$ denotes a solar mass.
These values significantly exceed the typical mass of a galaxy.
For example, the Andromeda galaxy possesses $M\sim10^{12}M_\odot$ \cite{kain2010}, and the Milky Way is estimated at $M\simeq0.5-2.0\times10^{12}M_\odot$ \cite{hayashi2021}.
However, the substantial total mass facilitates the exploration of BEC oscillations because $R_\mathrm{eq}$ hardly changes even if the total mass is reduced (see Fig \ref{figure1}).
For a BEC with the total mass of a typical galaxy, namely $M\sim10^{10}-10^{12}M_\odot$, the time scale of the breathing mode would extend to $10-100$ times greater than that presented in this paper, based on Eq. (\ref{periodicity_breathing}).
We set the parameters of the initial phase to $\tilde{\alpha}=\tilde{\beta}=\pm 0.03, 0.1$ to give an isotropic velocity field to a BEC in equilibrium. 

\begin{figure*}
\centering
\includegraphics[width=15.5cm]{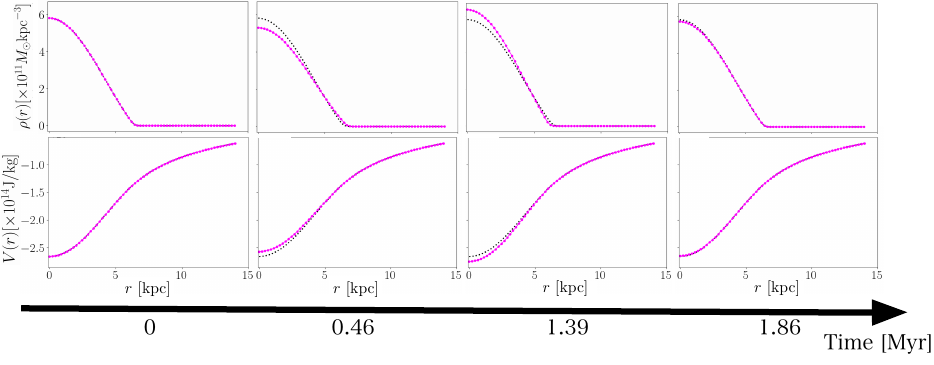}
\caption{
The density profile and the gravitational potential of the self-gravitating BEC.
The total mass is $M=2\times 10^{14}M_\odot$, and the initial phase is set by $\tilde{\alpha}=\tilde{\beta}=0.1$.
The radial coordinate, representing the distance from the center of our numerical domain, is depicted on the horizontal axis.
Each column portrays either the density profile or the gravitational potential at each time: $t=0, 0.46, 1.39, 1.86 ~\mathrm{Myr}$, where $\mathrm{Myr}$ means a megayear.
The magenta points show the data at each time, while the black dashed line shows the initial data.
The upper row displays the density profile, with the density depicted on the vertical axis, while the lower row exhibits the gravitational potential, with its value presented on the vertical axis.
}
\label{figure2}
\end{figure*}
Initially, we describe the time evolution of the density and gravitational potential profiles.
Despite solving the fully three-dimensional GPP equations in the simulations presented in this section, the configurations of the BECs maintain spherical symmetry throughout the time evolution.
Then, we consider the average of the density and gravitational potential over the solid angle to describe the time evolution, expressed as
\be
\rho(r,t)
=
\f{1}{4\pi r^2\Delta r}
\int_{r-\f{\Delta r}{2}}^{r+\f{\Delta r}{2}}dr'r'^2
\int d\Omega
\rho(\B{r'},t)
\ee
and
\be
V(r,t)
=
\f{1}{4\pi r^2\Delta r}
\int_{r-\f{\Delta r}{2}}^{r+\f{\Delta r}{2}}dr'r'^2
\int d\Omega
V(\B{r'},t),
\ee
where $\Omega$ denotes a solid angle and $\Delta r\approx0.22~\mathrm{kpc}$ is set so that $\Delta \tilde{r}$ matches the spatial resolution, namely $\Delta \tilde{r}=\tilde{L}_x/\tilde{N}_x$=0.3125.
Fig. \ref{figure2} shows the time evolution of the density and gravitational potential profiles.
The BEC periodically undergoes isotropic expansion and shrinkage.
Initially, the BEC expands by an outwardly directed initial velocity field, and the central density decreases until $t=0.46~\mathrm{Myr}$.
Subsequently, the BEC shrinks, and the central density increases until $t=1.39~\mathrm{Myr}$.
Finally, the BEC returns to its initial state at $t=1.86~\mathrm{Myr}$.
The gravitational potential also undergoes simultaneous oscillations with the density profile.
The gravitational potential becomes shallower as the central density decreases, and vice versa.
Note that at the outskirts, namely for $r\gtrsim R_\mathrm{TF}\approx6.46~\mathrm{kpc}$, the gravitational potential remains unchanged throughout the time evolution, maintaining $V=-GM/r$ due to the constancy of the total mass.

\begin{figure*}
\centering
\includegraphics[width=16cm]{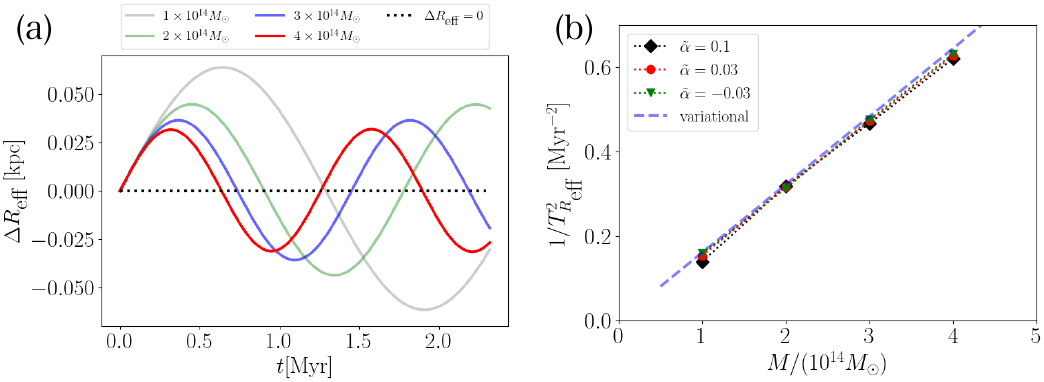}
\caption{
Oscillation of the effective radius $R_\mathrm{eff}(t)$ for each total mass of the self-gravitating BEC.
(a) The deviation of the effective radius from the initial/equilibrium one $\Delta R_\mathrm{eff}(t)=R_\mathrm{eff}(t)-R_\mathrm{eff}(0)$ when the initial phase is $\tilde{\alpha}=0.03$.
The horizontal axis shows the time, and the vertical axis shows $\Delta R_\mathrm{eff}(t)$.
Each solid curve is the time evolution of $\Delta R_\mathrm{eff}$ for different total masses (red: $4\times 10^{14}M_\odot$, blue: $3\times 10^{14}M_\odot$, green: $2\times 10^{14}M_\odot$ and black: $1\times 10^{14}M_\odot$).
The black dotted line shows the oscillation criterion, namely $\Delta R_\mathrm{eff}=0$.
(b) Total mass dependence on the period of the effective radius $T_{R_\mathrm{eff}}$ for various values of $\tilde{\alpha}$.
The horizontal axis shows the total mass, and the vertical axis shows the negative squared period $1/T_{R_\mathrm{eff}}^2$.
Each point shows the numerical result (black square: $\tilde{\alpha}=0.1$, red solid circle: $\tilde{\alpha}=0.03$ and green triangle: $\tilde{\alpha}=-0.03$).
The blue dashed line shows the analytical result obtained from the variational method, namely the negative squared period of breathing mode $1/T_\mathrm{B}^2$.
}
\label{figure3}
\end{figure*}
To quantitatively observe the spherical oscillation, we define the effective radius $R_\mathrm{eff}$ as
\be
R_\mathrm{eff}(t)
=
\sqrt{
\f{1}{M}\int d\B{r}
r^2\rho(\B{r},t)
}.
\label{eff_rad}
\ee
Using Eq. (\ref{trial_func}), it is proportional to the radius $R(t)$.
Hence, during the occurrence of the breathing mode in the self-gravitating BEC, the effective radius exhibits an oscillation similar to that of the radius.
Fig. \ref{figure3}(a) shows the deviation of the effective radius from its initial value, $\Delta R_\mathrm{eff}(t)=R_\mathrm{eff}(t)-R_\mathrm{eff}(0)$, for $\tilde{\alpha}=0.03$.
The graph shows sinusoidal curves for each total mass, where the oscillation period $T_{R_\mathrm{eff}}$ decreases as the total mass $M$ is increased.
Fig. \ref{figure3}(b) suggests that our results of $T_{R_\mathrm{eff}}$ agree quantitatively with Eq. (\ref{periodicity_breathing}), even for three values of $\tilde{\alpha}$.
Therefore, we conclude that the self-gravitating BEC induces the breathing mode.

\section{APPEARANCE OF ANISOTROPIC MODE}

\subsection{NUMERICAL RESULTS}

We contemplate the axisymmetric collective mode of a self-gravitating BEC.
Its total mass is $M/(10^{14}M_\odot)=1,2,3,4$, same as that in breathing mode.
These situations can be simulated by configuring $(\tilde{\alpha},\tilde{\beta})=(\pm0.03,0),~(0,\pm0.03)$.
Such parameters introduce an axisymmetric initial velocity field to the BEC in equilibrium and can cause anisotropic oscillation.

\begin{figure*}
\centering
\includegraphics[width=16cm]{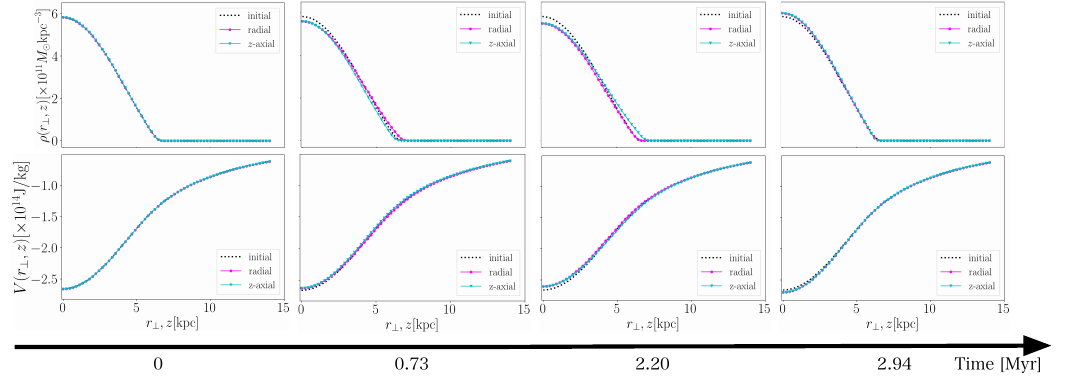}     
\caption{
The time evolution of the density profile $\rho(r_\perp,z)$ and the gravitational potential $V(r_\perp,z)$.
We consider the BEC has a total mass $M=2\times 10^{14}M_\odot$ and is endowed with an initial velocity by $\tilde{\alpha}=0.1,\tilde{\beta}=0$.
The horizontal axis shows $r_\perp$ and $z$.
Magenta points show the profiles at $z=0$, $\rho(r_\perp,z=0)$ and $V(r_\perp,z=0)$, while cyan triangles show those at $r_\perp=0$, $\rho(r_\perp=0,z)$ and $V(r_\perp=0,z)$.
The black dashed lines show the initial profile.
Each column shows either the density profile or the gravitational potential at each time: $t=0, 0.73, 2.20, 2.94 ~\mathrm{Myr}$, where Myr means a megayear.
The upper row shows the density profile, and the vertical axis shows the density.    
The lower row shows the gravitational potential, and the vertical axis shows its value.
}
\label{figure4}
\end{figure*}
To describe the time evolution of the BEC, we obtain the density and gravitational potential profiles by averaging $\rho(\B{r},t)$ and $V(\B{r},t)$:
\be
\rho(r_\perp,z,t)
=
\f{1}{2\pi r_\perp\Delta r_\perp}
\int_{r_\perp-\f{\Delta r_\perp}{2}}^{r_\perp+\f{\Delta r_\perp}{2}}dr'_\perp r'_\perp
\int_0^{2\pi}d\theta
\rho(r'_\perp,\theta,z,t)
\label{density_profile_cy}
\ee
and
\be
V(r_\perp,z,t)
=
\f{1}{2\pi r_\perp\Delta r_\perp}
\int_{r_\perp-\f{\Delta r_\perp}{2}}^{r_\perp+\f{\Delta r_\perp}{2}}dr'_\perp r'_\perp
\int_0^{2\pi}d\theta
V(r'_\perp,\theta,z,t),
\label{potential_profile_cy}
\ee
where $\Delta r_\perp \approx0.22~\mathrm{kpc}$ is determined in the same manner as $\Delta r$, namely $\Delta \tilde{r}_\perp=\tilde{L}_x/\tilde{N}_x=0.3125$, and $\theta=\tan^{-1}(y/x)$.
Fig.\ref{figure4} shows the density profile and the gravitational potential.
They show synchronous oscillations similar to those in the breathing mode.
However, $\rho(r_\perp,z=0,t)$ is different from $\rho(r_\perp=0,z,t)$.
At $t=0.73~\mathrm{Myr}$, the former shows BEC expansion, while the latter shows shrinkage.
Subsequently, the density profile at $t=2.20~\mathrm{Myr}$ exhibits an inverse behavior: a decrease in $\rho(r_\perp,z=0,t)$ and an increase in $\rho(r_\perp=0,z,t)$.
By $t=2.94~\mathrm{Myr}$, the density profile reverts to a spherical shape similar to the initial profile.
Hence, the self-gravitating BEC induces anisotropic oscillations due to the nonspherical initial velocity field.
We observe this behavior for other combinations of $(\tilde{\alpha},\tilde{\beta})$ such as $\tilde{\alpha}\neq\tilde{\beta}$  (e.g. $(\tilde{\alpha},\tilde{\beta})=(\pm0.1,\mp0.03)$).
Note that the gravitational potential remains in the shape of $1/r$ at the outskirts throughout our simulations although the density profile anisotropically deforms.

\begin{figure*}
\centering
\includegraphics[width=16cm]{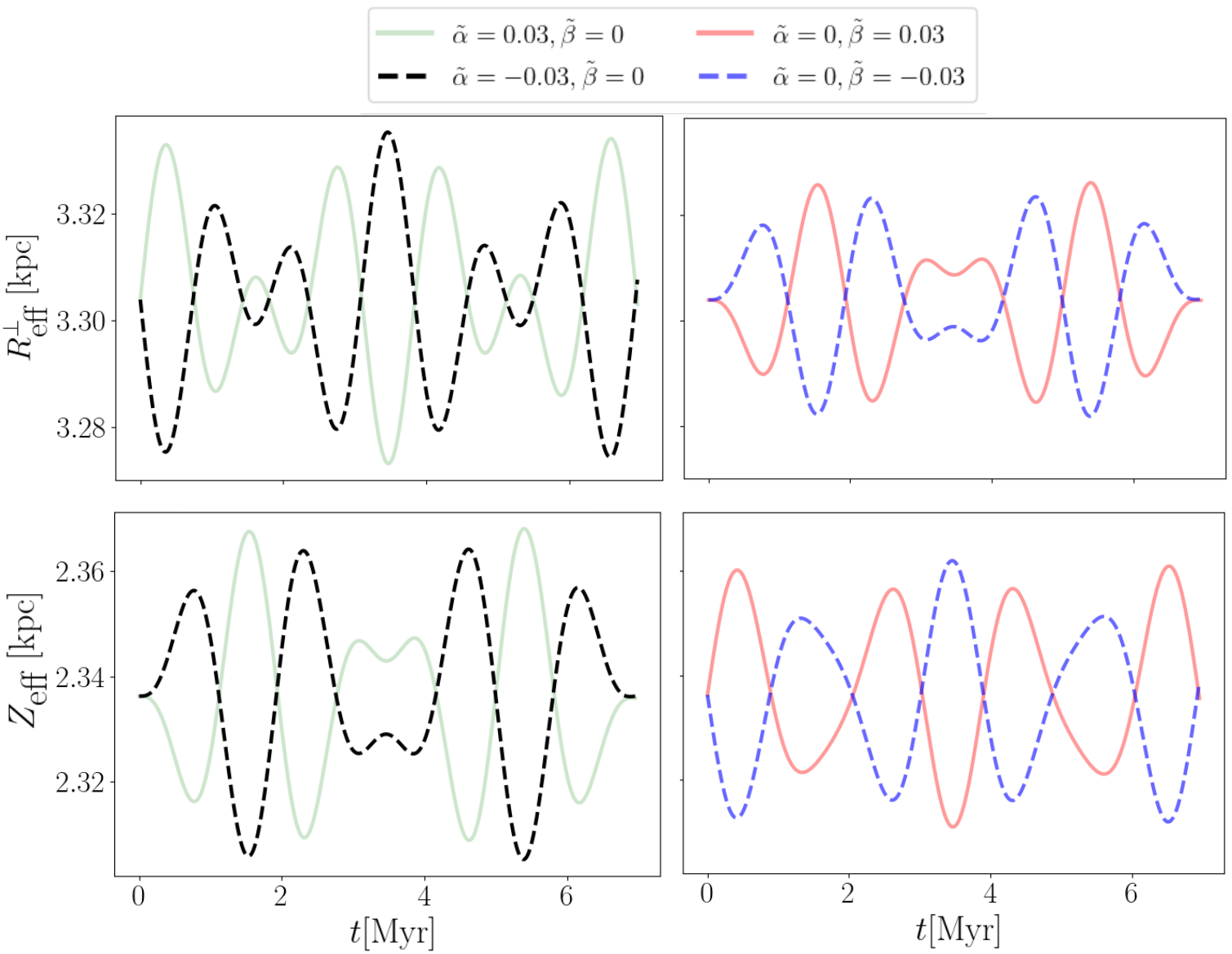}
\caption{
Oscillations of the effective width $R_\mathrm{eff}^\perp,~Z_\mathrm{eff}$.
The self-gravitating BEC possesses a total mass $M=4\times 10^{14}M_\odot$, and the initial velocity is set by $(\tilde{\alpha},~\tilde{\beta})=(\pm0.03,0),~(0,\pm0.03)$.
The horizontal axis shows the time.
The left column shows the result of $(\tilde{\alpha},~\tilde{\beta})=(\pm0.03,0)$, which means the initial velocity in the direction of $r_\perp$.
The green solid line shows the oscillation when the BEC initially expands ($\tilde{\alpha}>0$), and the black dashed line shows the oscillation when the BEC initially shrinks ($\tilde{\alpha}<0$).
Conversely, the right column shows the result of $(\tilde{\alpha},~\tilde{\beta})=(0,\pm0.03)$, indicating the initial velocity along the z-axis.
The red solid line shows the oscillation when the BEC initially expands ($\tilde{\beta}>0$), and the blue dashed line shows the BEC initially shrinks ($\tilde{\beta}<0$).
While the upper row shows the time evolution of $R_\mathrm{eff}^\perp$, the lower row shows the time evolution of $Z_\mathrm{eff}$.
}
\label{figure5}
\end{figure*}
To extract the effective degrees of freedom from the anisotropic oscillation, we introduce two quantities $R_\mathrm{eff}^\perp$ and $Z_\mathrm{eff}$ defined as
\be
R_\mathrm{eff}^\perp(t)
=
\sqrt{
\f{1}{M}\int d\B{r}
r_\perp^2\rho(\B{r},t)
},
\label{r_width}
\ee
\be
Z_\mathrm{eff}(t)
=
\sqrt{
\f{1}{M}\int d\B{r}
z^2\rho(\B{r},t)
}.
\label{z_width}
\ee
These are the effective widths of the BEC in each direction; $R_\mathrm{eff}^\perp$ characterizes the width in the direction of the radial coordinate $r_\perp$ and $Z_\mathrm{eff}$ is the width along the $z$-axis.
Fig.\ref{figure5} shows the complicated oscillatory behavior of these quantities, describing the opposing oscillations between $R_\mathrm{eff}^\perp$ and $Z_\mathrm{eff}$.
The BEC exhibits shrinkage along the $z$-axis when a positive velocity is initially added in the direction of $r_\perp$, and vice versa.
Similarly, the BEC shows shrinkage in the direction of $r_\perp$ in response to a positive initial velocity in the direction of the $z$-axis, and vice versa.

\begin{figure}
\centering
\includegraphics[width=8.5cm]{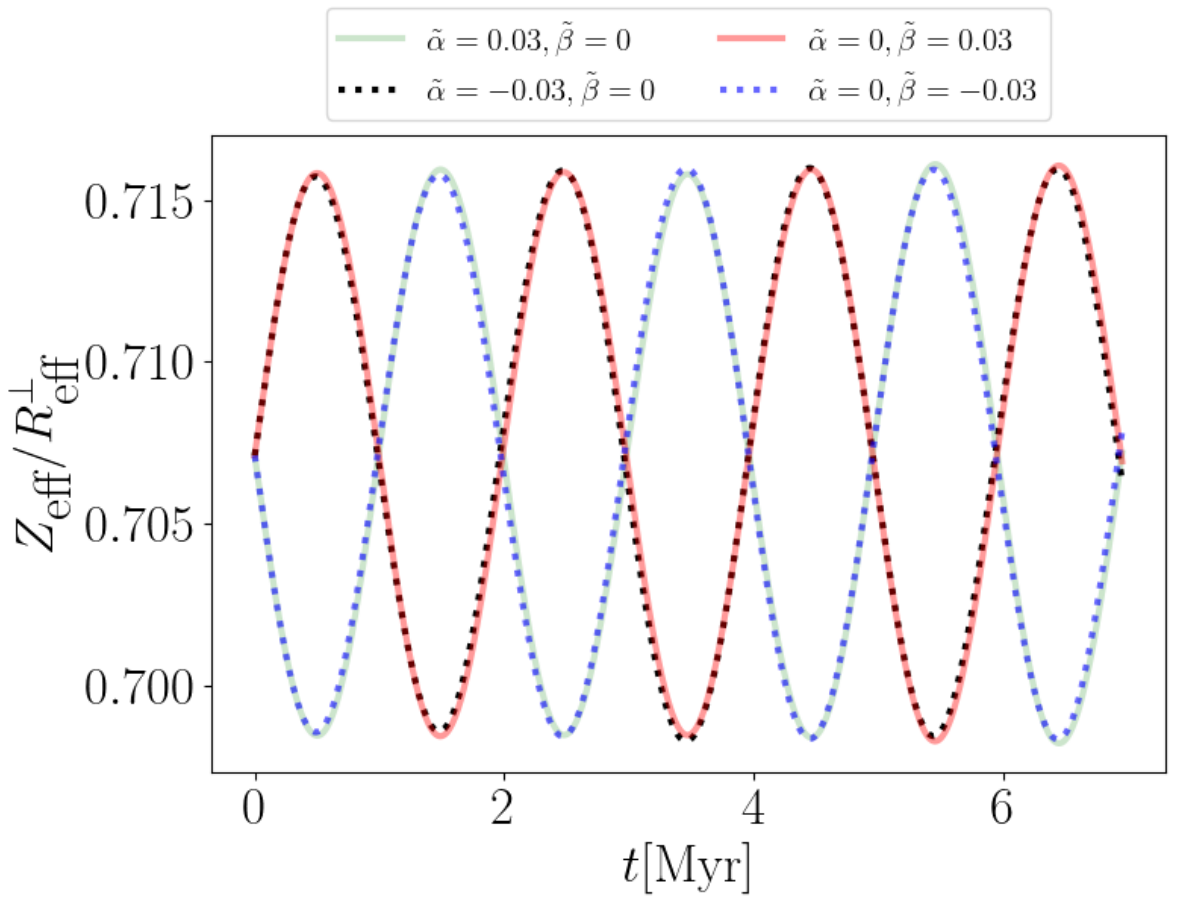}
\caption{
The oscillation of the ratio $Z_\mathrm{eff}/R_\mathrm{eff}^\perp$ when the total mass of the self-gravitating BEC is $M=4\times 10^{14}M_\odot$.
The horizontal axis shows the time, and the vertical axis shows the value of the ratio $Z_\mathrm{eff}/R_\mathrm{eff}^\perp$.
The solid lines show cases of initial expansion (green: the direction of $r_\perp$ and red: along the $z$-axis), whereas the dotted lines exhibit cases of initial shrinkage (black: the direction of $r_\perp$ and blue: the direction of the $z$-axis).
}
\label{figure6}
\end{figure}
Although $R_\mathrm{eff}^\perp$ and $Z_\mathrm{eff}$ manifest intricate oscillations, a harmonic oscillation can be obtained from the ratio $Z_\mathrm{eff}/R_\mathrm{eff}^\perp$, as shown in Fig. \ref{figure6}.
Among the four cases, the oscillations have a common amplitude and frequency despite the different behaviors of $R_\mathrm{eff}^\perp,~Z_\mathrm{eff}$.
The ratio $Z_\mathrm{eff}/R_\mathrm{eff}^\perp$ indicates the extent to which the BEC axisymmetrically deforms from a spherical shape, similar to the aspect ratio of the BEC.
It can be inferred that one of the coupled modes is the quadrupole mode because the harmonic mode of the aspect ratio suggests a quadrupole mode of the BEC \cite{Jin1996, Mewes1996, Deppner2021}.

\begin{figure}
\centering
\includegraphics[width=8.5cm]{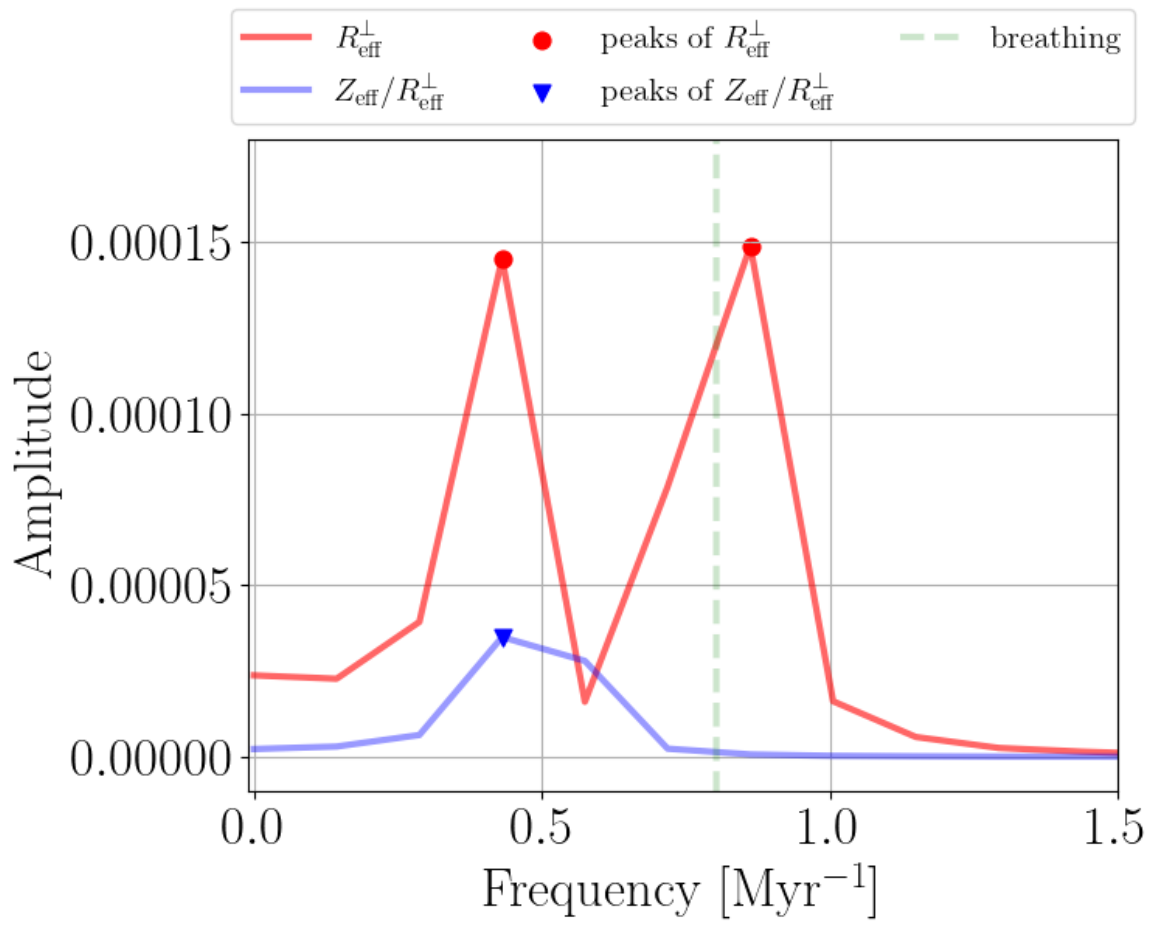}
\caption{
The Fourier transformation of $R_\mathrm{eff}^\perp$ and $Z_\mathrm{eff}/R_\mathrm{eff}^\perp$.
The self-gravitating BEC has a total mass $M=4\times 10^{14}M_\odot$, with an the initial velocity given by $\tilde{\alpha}=0.03,~\tilde{\beta}=0$.
The horizontal axis shows the frequency, while the vertical axis shows the amplitude.
The solid lines show the results of each transformation (red: $R_\mathrm{eff}^\perp$ and blue: $Z_\mathrm{eff}/R_\mathrm{eff}^\perp$), and the green dashed line shows the analytical/variational result of the breathing mode $T_B$.
The red points show the peaks of $R_\mathrm{eff}^\perp$, and the blue triangle shows the peak of $Z_\mathrm{eff}/R_\mathrm{eff}^\perp$.
}
\label{figure7}
\end{figure}
The Fourier transformation serves as a valuable tool for decomposing complicated oscillations into eigenmodes.
Fig. \ref{figure7} shows the Fourier transformation of $R_\mathrm{eff}^\perp$ and $Z_\mathrm{eff}/R_\mathrm{eff}^\perp$, as shown in Fig. \ref{figure5} and Fig. \ref{figure6}.
Within this representation, two peaks of $R_\mathrm{eff}^\perp$ and a peak of $Z_\mathrm{eff}/R_\mathrm{eff}^\perp$ can be discerned.
The high-frequency peak of $R_\mathrm{eff}^\perp$ corresponds to the period of the breathing mode $T_B$ with the corresponding total mass.
On the other hand, the low-frequency peak in $R_\mathrm{eff}^\perp$ closely matches that of $Z_\mathrm{eff}/R_\mathrm{eff}^\perp$.
This peak suggests the frequency associated with the quadrupole mode of the BEC.
Hence, the self-gravitating BEC causes an anisotropic collective mode wherein the quadruple mode superposes the breathing mode due to the axisymmetric initial velocity field.

\begin{figure*}
\centering
\includegraphics[width=16cm]{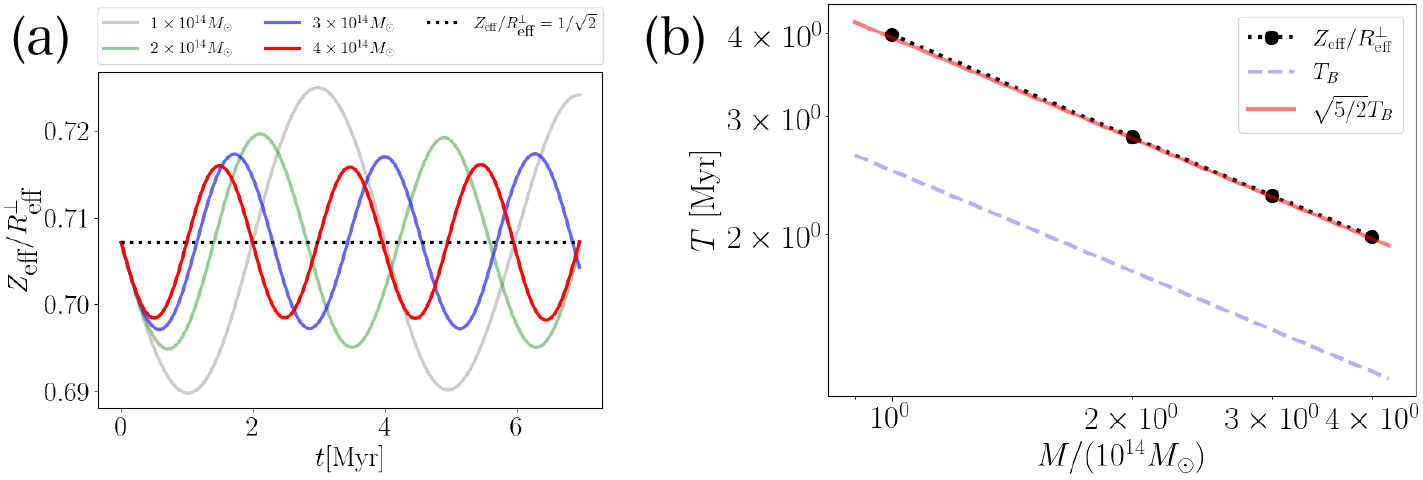}
\caption{
The oscillation of $Z_\mathrm{eff}/R_\mathrm{eff}^\perp$ for each total mass.
The initial phase is determined by $\tilde{\alpha}=0.03, \tilde{\beta}=0$.
(a) The time evolution of $Z_\mathrm{eff}/R_\mathrm{eff}^\perp$ for each total mass.
The horizontal axis shows time, while the vertical axis shows the value of $Z_\mathrm{eff}/R_\mathrm{eff}^\perp$.
The black dotted line shows the ratio when the BEC is in equilibrium; the ratio becomes $Z_\mathrm{eff}/R_\mathrm{eff}^\perp=1/\sqrt{2}$.
Each solid line shows the result (red: $4\times10^{14}M_\odot$, blue: $3\times10^{14}M_\odot$, green: $2\times10^{14}M_\odot$, black: $1\times10^{14}M_\odot$).
(b) The dependence of the period of $Z_\mathrm{eff}/R_\mathrm{eff}^\perp$ on the total mass by the log-log plot.
The horizontal axis shows the total mass, while the vertical axis shows the period.
The black points show the numerical results for each total mass, whereas the blue dashed line shows the variational result of the breathing mode $T_B$.
The red solid line shows $\sqrt{5/2}T_B$.
}
\label{figure8}
\end{figure*}
We vary the parameters of the initial phase and total mass to investigate the dependence of the period of $Z_\mathrm{eff}/R_\mathrm{eff}^\perp$ on these quantities.
Although we study the dynamics of various combinations of $\tilde{\alpha}$ and $\tilde{\beta}$, the period remains constant for all the simulations.
Thus, the period of $Z_\mathrm{eff}/R_\mathrm{eff}^\perp$ is independent of the initial velocity field.
However, the total mass can change the period of $Z_\mathrm{eff}/R_\mathrm{eff}^\perp$, as shown in Fig. \ref{figure8}.
Fig. \ref{figure8} (a) shows that $Z_\mathrm{eff}/R_\mathrm{eff}^\perp$ exhibits harmonic oscillation similar to $R_\mathrm{eff}$ of the breathing mode induced by the isotropic initial velocity field; the period becomes short as the total mass increases.
The $M$-dependence of the period of $Z_\mathrm{eff}/R_\mathrm{eff}^\perp$ and its comparison with $T_B$ are shown in Fig. \ref{figure8} (b).
We find that $Z_\mathrm{eff}/R_\mathrm{eff}^\perp$ shows a straight line parallel to $T_B$ in the log-log plot, indicating that the period of the quadrupole mode is also inversely proportional to $\sqrt{M}$, similar to the breathing mode.
However, the period of $Z_\mathrm{eff}/R_\mathrm{eff}^\perp$ is approximately $1.57$ times larger than $T_B$; in other words, the quadrupole mode exhibits lower frequency oscillation compared to the breathing mode.
This characteristic can be also observed in atomic BECs confined by an isotropic harmonic potential. 
The frequency of the breathing mode is $\sqrt{5/2}$ times that of the quadrupole mode \cite{stringari}.
Hence, our numerical results are consistent with those of typical atomic BECs trapped by external potentials.
The ratio of the period of $Z_\mathrm{eff}/R_\mathrm{eff}^\perp$ to $T_B$ closely resembles that of the BEC confined by an isotropic harmonic potential.
Moreover, the periods of $Z_\mathrm{eff}/R_\mathrm{eff}^\perp$ are in good agreement with $\sqrt{5/2}T_B$ (see Fig. \ref{figure8}).
Thus, the numerical results indicate that the quadrupole mode has the period $\sqrt{5/2}T_B$.

The period of the quadrupole mode in a self-gravitating BEC can also be analytically confirmed by a sum-rule approach \cite{Giovanazzi2001}.
This approach, based on the linear response theory, enables us to derive the frequency of the collective mode without explicitly solving the equation of motion \cite{stringari}.
Applying the approach to the self-gravitating BEC, the angular frequency of the quadrupole mode $\Omega_Q$ is generally expressed as
\be
\Omega_Q^2
=
\f{4K-\f{4}{5}W}{M\langle r^2\rangle},
\label{quadrupole_sumrule}
\ee
where $\langle r^2\rangle=\int d\B{r}r^2\rho(\B{r})/M$.
Particularly, within the TF region, the period $T_Q=2\pi/\Omega_Q$ can be approximated as
\be
T_Q
\simeq
2\pi
\sqrt{
\f{5}{4}\f{M\langle r^2\rangle}{-W}
}
=
2\pi
\sqrt{
\f{5\pi(\pi^2-6)}{3}
\sqrt{\f{\hbar^{6}a^{3}}{G^{5}m^{9}}}
}
\f{1}{\sqrt{M}}.
\label{quadrupole_sumrule_TF}
\ee
Here, we use Eqs. (\ref{tf}) and (\ref{potential_eq}) to obtain Eq. (\ref{quadrupole_sumrule_TF}).
\begin{figure*}
\centering
\includegraphics[width=9cm]{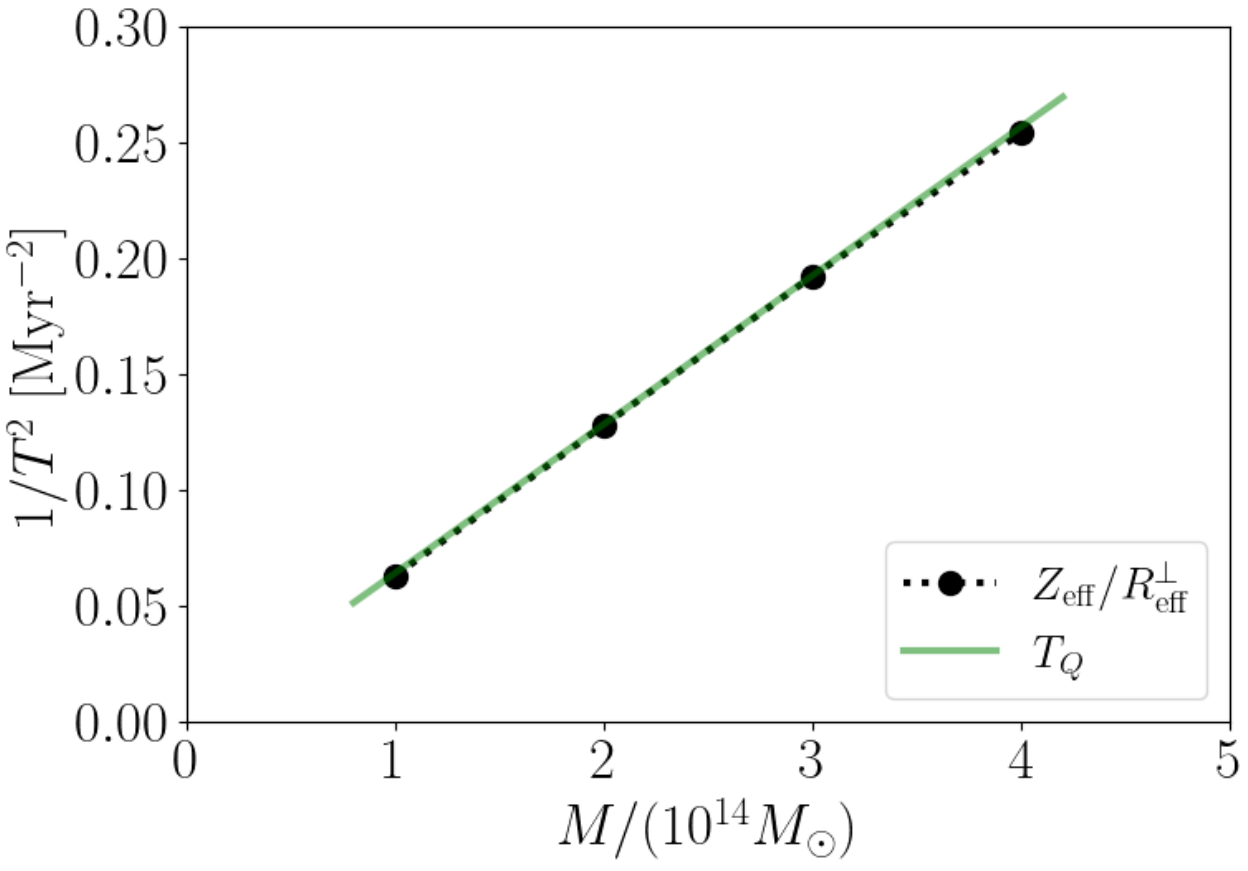}
\caption{
The comparison of the period of $Z_\mathrm{eff}/R_\mathrm{eff}^\perp$ to that from a sum-rule approach.
The horizontal axis shows the total mass, while the vertical axis shows the negative squared period.
The black points show the numerical results for each total mass, whereas the green solid line shows the analytical result written by Eq. (\ref{quadrupole_sumrule_TF}).
}
\label{figure10}
\end{figure*}
In Fig. \ref{figure10}, the period of $Z_\mathrm{eff}/R_\mathrm{eff}^\perp$ is compared with Eq. (\ref{quadrupole_sumrule_TF}), yielding quantitative consistency.
This approach also yields the general angular frequency of the breathing mode $\Omega_B$ given by
\be
\Omega_B^2
=
\f{4K+W+9I}{M\langle r^2\rangle}.
\label{freq_b_sumrule}
\ee
Using Eqs. (\ref{quadrupole_sumrule}), (\ref{freq_b_sumrule}), and the virial theorem, $2K+W+3I=0$, the ratio $\Omega_B/\Omega_Q$ in the TF regime becomes
\be
\f{\Omega_B}{\Omega_Q}
=
\sqrt{
\f{5}{2}
\f{3+(K/I)}{3+7(K/I)}
}
\simeq
\sqrt{\f{5}{2}},
\label{}
\ee
which agrees with our result.
Hence, our numerical result regarding the periods of the quadrupole modes is quantitatively consistent with the previous study \cite{Giovanazzi2001} using a sum-rule approach.

To reproduce our numerical results in more detail, we extend the variational method from the spherical self-gravitating BEC in Sec.$\mathrm{I}\hspace{-1.2pt}\mathrm{I}\hspace{-1.2pt}\mathrm{I}$-A to the axisymmetric system in the subsequent subsection.

\subsection{ANALYSIS OF AXISYMMETRIC COLLECTIVE MODE BY VARIATIONAL METHOD}

Rindler-Daller and Shapiro extended the TF solution to an ellipsoidal self-gravitating BEC \cite{ACE,daller2012} based on the ellipsoidal approximation \cite{lai1993}. 
Given that the self-gravitating BEC takes a spheroidal configuration with semi-axes $R_\perp$ and $Z$, the extended TF solution is expressed as
\be
\rho(q)
=\f{\pi M}{4R_\perp^2Z}j_0\left( \f{\pi q}{R_\perp}\right),
\label{spheroid_tf}
\ee
where $M$ denotes the total mass and $q\in(0,R_\perp]$ satisfies $(q/R_\perp)^2=(r_\perp/R_\perp)^2+(z/Z)^2$.
Employing Eq. (\ref{spheroid_tf}), the trial function is set as
\be
    \psi(\B{r},t)
    =
    \sqrt{\f{\pi M}{4mR_\perp(t)^2Z(t)}j_0\left(\f{\pi q}{R_\perp(t)}\right)}
    \exp\left[i\f{m}{2\hbar}\left(r_\perp^2H_\perp(t)+z^2H_z(t)\right)\right],
\label{aniso_trial_func}
\ee
where $H_\perp(t)$ and $H_z(t)$ are variables that provide the velocity field such that 
\be
\B{v}=\left(r_\perp H_\perp(t)\right)\B{e}_\perp+\left(zH_z(t)\right)\B{e}_z.
\label{vel_rz}
\ee

The Lagrangian can be derived by substituting Eq. (\ref{aniso_trial_func}) into its definition (\ref{lagrangian_gpp}).
Using Eq. (\ref{aniso_trial_func}), the energy components $K$, $W$, $I$ defined by Eqs. (\ref{kinetic_energy}), (\ref{potential_energy}), and (\ref{self-interaction_energy}) are expressed as
\begin{align}
\label{kinetic_eq_spheroid}
&
    \begin{aligned}
    K
    \simeq
    &\f{\pi}{24}\f{\hbar^2M}{m^2}F(\Lambda)\left(\f{2}{R_\perp(t)^2}+\f{1}{Z(t)^2}\right) \\
    &+\f{M}{6}\f{\pi^2-6}{\pi^2}\left\{2H_\perp(t)^2R_\perp(t)^2+H_z(t)^2Z(t)^2\right\},
    \end{aligned}
\\
\label{potential_eq_spheroid}
&W
=
-\f{3GM^2}{4}\f{1}{\{R_\perp(t)^2Z(t)\}^{1/3}}, \\
\label{self-interaction_eq_spheroid}
&I
=
\f{\pi^2}{4}\f{\hbar^2aM^2}{m^3}\f{1}{R_\perp(t)^2Z(t)},
\end{align}
where Eq. (\ref{potential_eq_spheroid}) can be derived using the formula for the gravitational energy of the spheroidal density profile in \cite{Binney}. 
We assume that the deformation of the BEC is sufficiently small\footnote{In general, the gravitational energy of a spheroid is affected by its deformation (see \cite{Binney}). However, we neglect these effects because, in the present study, we address only small oscillations.}.
Hence, the Lagrangian can be written as
\be
    L(R_\perp,Z,H_\perp,H_z) \\
    =
    -\f{\pi^2-6}{6\pi^2}M
    \left\{
    2R_\perp^2(\dot{H}_\perp+H_\perp^2)
    +Z^2(\dot{H}_z+H_z^2)
    \right\} \\
    -U(R_\perp,Z).
\label{aniso_lagranian}
\ee
Here, the effective potential $U(R_\perp,Z)$ is defined as
\be
    U(R_\perp,Z)
    =
    \f{C_z}{3}\left(\f{2}{R_\perp^2}+\f{1}{Z^2}\right)-\f{C_p}{(R_\perp^2Z)^{1/3}}+\f{C_i}{R_\perp^2Z},
\label{aniso_eff_pot}
\ee
where $C_z$, $C_p$, and $C_i$ are identical to those in Eq. (\ref{eff_pot}).
The Euler-Lagrange equations for $H_\perp(t)$ and $H_z(t)$ are $H_\perp(t)=\dot{R}_\perp(t)/R_\perp(t)$ and $H_z(t)=\dot{Z}(t)/Z(t)$.
Therefore, the Euler-Lagrange equations for $R_\perp(t)$ and $Z(t)$ are
\begin{empheq}[left=\empheqlbrace]{align}
\label{el_r}
\f{2}{3}\f{\pi^2-6}{\pi^2}M\f{d^2R_\perp(t)}{dt^2}=-\f{\p U(R_\perp,Z)}{\p R_\perp},
\\
\label{el_z}
\f{1}{3}\f{\pi^2-6}{\pi^2}M\f{d^2Z(t)}{dt^2}=-\f{\p U(R_\perp,Z)}{\p Z}.
\end{empheq}
These Euler-Lagrange equations show the spherical equilibrium state with $R_\perp=Z=R_\mathrm{eq}$, which is consistent with the results in Sec.$\mathrm{I}\hspace{-1.2pt}\mathrm{I}\hspace{-1.2pt}\mathrm{I}$-A.

Similarly to the spherically symmetric case, we consider a small oscillation of the semi-axes near $R_\mathrm{eq}$.
The fluctuations in the semi-axes are given by $R_\perp(t)=R_\mathrm{eq}+\delta R_\perp(t)$ and $Z(t)=R_\mathrm{eq}+\delta Z(t)$ ($|\delta R_\perp(t)|,~|\delta Z(t)|\ll R_\mathrm{eq}$).
The equations of motion for the fluctuations $\delta R_\perp(t)$ and $\delta Z(t)$ are given by
\be
\f{3(\pi^2-6)}{\pi^2}MR_\mathrm{eq}^5\f{d^2}{dt^2}
\begin{pmatrix}
   2\delta R_\perp(t) \\
   \delta Z(t)
\end{pmatrix}
=
-
\begin{pmatrix}
  4C_pR_\mathrm{eq}^2 & 2(-C_pR_\mathrm{eq}^2+9C_i) \\
  -C_pR_\mathrm{eq}^2+9C_i & 5C_pR_\mathrm{eq}^2-9C_i
\end{pmatrix}
\begin{pmatrix}
   2\delta R_\perp(t) \\
   \delta Z(t)
\end{pmatrix}
.
\label{eom_rz}
\ee
When we write the solutions as $2\delta R_\perp(t)=A\exp[i\omega t]$ and $\delta Z(t)=B\exp[i\omega t]$, the angular frequency $\omega$ satisfies
\be
\omega
=
\pm\sqrt{
\f{2\pi^2}{\pi^2-6}\f{C_zR_\mathrm{eq}+3C_i}{MR_\mathrm{eq}^5}
}
\equiv
\pm\omega_B,
\label{frequency_b}
\ee
or
\be
\omega
=
\pm 2
\sqrt{
\f{\pi^2}{\pi^2-6}\f{C_z}{MR_\mathrm{eq}^4}
}
\equiv
\pm\omega_Q
.
\label{frequency_q}
\ee
Eq. (\ref{frequency_b}) provides the eigenfrequency of the breathing mode compared to the coefficient of Eq. (\ref{osc_eom_g}).
Indeed, when $\omega=\pm\omega_B$, the eigenmodes satisfy $A=2B$, \textit{i.e.}
\be
\begin{pmatrix}
   \delta R_\perp(t) \\
   \delta Z(t)
\end{pmatrix}
=
A
\begin{pmatrix}
   1 \\
   1
\end{pmatrix}
\exp\left[\pm i\omega_B t\right].
\label{b_mode}
\ee
This indicates that the BEC either undergoes spherical expansion or shrinkage.
Conversely, when $\omega=\pm\omega_Q$, the eigenmodes satisfy $A+B=0$, \textit{i.e.} 
\be
\begin{pmatrix}
   \delta R_\perp(t) \\
   \delta Z(t)
\end{pmatrix}
=
A
\begin{pmatrix}
   1 \\
   -2
\end{pmatrix}
\exp\left[\pm i\omega_Q t\right].
\label{q_mode}
\ee
We can consider Eq. (\ref{q_mode}) as the quadrupole mode because it shows that the BEC elongates along the $z$-axis or expands perpendicular to the $z$-axis.

We compare Eqs. (\ref{b_mode}) and (\ref{q_mode}) with the numerical results.
In our numerical simulations, we establish two initial conditions: $\delta R_\perp(t=0)=\delta Z(t=0)=0$ and $\B{\tilde{v}}(\tilde{t}=0,\tilde{r}_\perp,\tilde{z})=\tilde{\alpha}\tilde{r}_\perp\B{e}_\perp+\tilde{\beta}\tilde{z}\B{e}_z$.
As a result, the fluctuations $\delta R_\perp(t)$ and $\delta Z(t)$ become
Consequently, the perturbations $\delta R_\perp(t)$ and $\delta Z(t)$ are given by
\be
    \delta R_\perp(t)
    =
    \f{\tilde{\lambda}^2c^2mR_\mathrm{eff}}{3\hbar}
    \left\{
    \f{2\tilde{\alpha}+\tilde{\beta}}{\omega_B}\sin\left(\omega_B t\right)
    +
    \f{\tilde{\alpha}-\tilde{\beta}}{\omega_Q}\sin\left(\omega_Q t\right)
    \right\},    
\label{dr_perp}
\ee
and
\be
    \delta Z(t)
    =
    \f{\tilde{\lambda}^2c^2mR_\mathrm{eff}}{3\hbar}
    \left\{
    \f{2\tilde{\alpha}+\tilde{\beta}}{\omega_B}\sin\left(\omega_B t\right)
    -
    2\f{\tilde{\alpha}-\tilde{\beta}}{\omega_Q}\sin\left(\omega_Q t\right)
    \right\}.
\label{dz}
\ee
The effective widths $R_\mathrm{eff}^\perp$ and $Z_\mathrm{eff}$ are respectively given by $R_\mathrm{eff}^\perp=\sqrt{2/3}\sqrt{(\pi^2-6)/\pi^2}R_\perp$ and $Z_\mathrm{eff}=\sqrt{(\pi^2-6)/(3\pi^2)}Z$ using Eq. (\ref{aniso_trial_func}).
Thus, we obtain
\be
\begin{split}
\f{Z_\mathrm{eff}}{R_\mathrm{eff}^\perp}
=
\f{Z}{\sqrt{2}R_\perp}
\simeq
\f{1}{\sqrt{2}}-\f{\tilde{\lambda}^2c^2m(\tilde{\alpha}-\tilde{\beta})}{\sqrt{2}\hbar\omega_Q}\sin\left(\omega_Q t\right),
\end{split}
\label{eff_asp}
\ee
which shows that the quadrupole mode induces a harmonic oscillation of $Z_\mathrm{eff}/R_\mathrm{eff}^\perp$ near $1/\sqrt{2}$.
This confirms that the quadrupole mode appears in the numerical simulations ( Fig. \ref{figure6} or Fig. \ref{figure8} (a) in Sec.$\mathrm{I}\hspace{-1.2pt}\mathrm{V}$-A ).

\begin{figure}
\centering
\includegraphics[width=8.5cm]{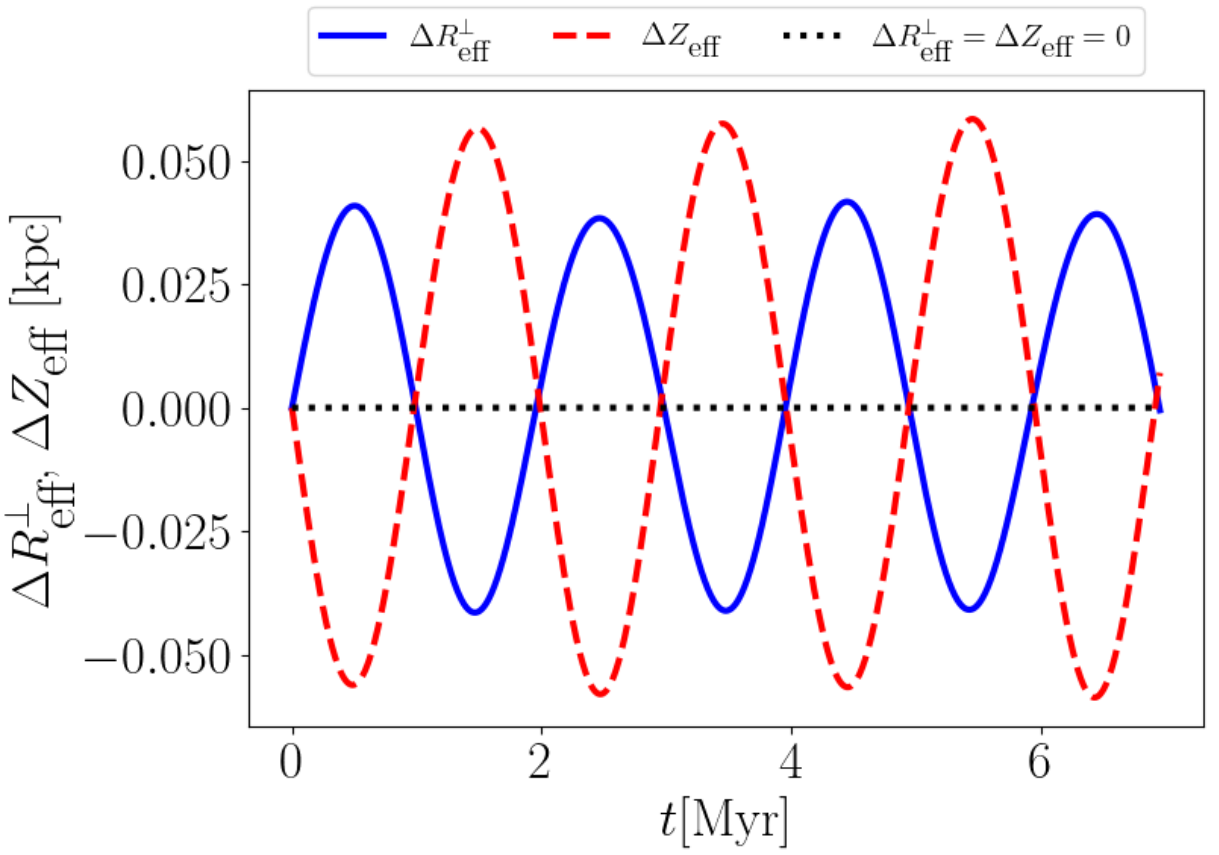}
\caption{
The variation in the effective widths $\Delta R_\mathrm{eff}^\perp=R_\mathrm{eff}^\perp(t)-R_\mathrm{eff}^\perp(0)$ and $\Delta Z_\mathrm{eff}=Z_\mathrm{eff}(t)-Z_\mathrm{eff}(0)$ when a total mass is $M=4\times10^{14}M_\odot$ and the initial phase is set by $\tilde{\alpha}=0.03$ and $\tilde{\beta}=-0.06$.
The horizontal axis shows time and the vertical axis shows $\Delta R_\mathrm{eff}^\perp$ and $\Delta Z_\mathrm{eff}$.
The blue solid line shows the time evolution of $\Delta R_\mathrm{eff}^\perp$, and the dashed red line shows the time evolution of $\Delta Z_\mathrm{eff}$.
The dotted black line shows that $\Delta R_\mathrm{eff}^\perp=\Delta Z_\mathrm{eff}=0$.
}
\label{figure9}
\end{figure}
Our variational calculation of the axisymmetric self-gravitating BEC suggests that solely the quadrupole mode can be extracted through the selection of an appropriate initial phase.
As the first term on the right-hand side of Eqs. (\ref{dr_perp}) and (\ref{dz}) vanish, the exclusive extraction of the quadrupole mode is feasible by setting the initial phase as $2\tilde{\alpha}+\tilde{\beta}=0$.
Consequently, the quadrupole mode shows 
\be
\delta R_\perp(t)
=
\f{\tilde{\alpha}\tilde{\lambda}^2c^2mR_\mathrm{eq}}{\hbar\omega_Q}
\sin\left(\omega_Q t\right)
\label{q_dr}
\ee
and
\be
\delta Z(t)
=
-2
\f{\tilde{\alpha}\tilde{\lambda}^2c^2mR_\mathrm{eq}}{\hbar\omega_Q}
\sin\left(\omega_Q t\right),
\label{q_dz}
\ee
resulting in harmonic oscillations of $R_\mathrm{eff}^\perp$ and $Z_\mathrm{eff}$ in opposite phase.
Indeed, we can numerically extract only the quadrupole mode when the total mass is $M=4\times10^{14}M_\odot$ and the initial phase is set as $\tilde{\alpha}=0.03$ and $\tilde{\beta}=-0.06$, as shown in Fig. \ref{figure9}.
This figure shows that $\Delta R_\mathrm{eff}^\perp\equiv R_\mathrm{eff}^\perp(t)-R_\mathrm{eff}^\perp(0)$ and $\Delta Z_\mathrm{eff}\equiv Z_\mathrm{eff}(t)-Z_\mathrm{eff}(0)$ oscillate monotonically in opposite phases, with periods identical to those shown in Fig. \ref{figure6}.
Additionally, the amplitude of $\Delta Z_\mathrm{eff}$ is approximately $1.40$ times larger than that of $\Delta R_\mathrm{eff}^\perp$, aligning with our analysis, as Eqs. (\ref{q_dr}) and (\ref{q_dz}) predict that the ratio of the amplitudes of $\Delta Z_\mathrm{eff}$ and $\Delta R_\mathrm{eff}^\perp$ is $\sqrt{2}$.
Therefore, we successfully extract only the quadrupole mode of self-gravitating BEC.

Although this variational method qualitatively agrees with the numerical results, the frequency or period of the quadrupole mode diverges from them.
The ratio of $\omega_Q$ to $\omega_B$ is given by
\be
\f{\omega_Q}{\omega_B}
=
2\sqrt{\f{C_zR_\mathrm{eq}}{C_pR_\mathrm{eq}^2+3C_i}},
\label{bq_ratio}
\ee
using Eqs. (\ref{mass_radius}), (\ref{frequency_b}) and (\ref{frequency_q}).
However, this ratio almost vanishes because we use $C_pR_\mathrm{eq}^2\simeq3C_i$ and $C_zR_\mathrm{eq}\ll C_i$ in the TF approximation.
This is different from the numerical results indicating $\omega_Q/\omega_B=\sqrt{2/5}$ (see Fig. \ref{figure8} (b)).
This inconsistency arises because the trial function in Eq. (\ref{aniso_trial_func}) is based on the TF solution, which neglects kinetic energy and becomes invalid near the surface of the BEC.
However, Eqs. (\ref{kinetic_eq}) and (\ref{frequency_q}) show that the kinetic energy of the equilibrium state predominantly influences the frequency of the quadrupole mode.
Notably, the quadrupole mode does not oscillate the central density, and neither does it involve potential nor self-interaction energy contributions.
Consequently, it only changes the density near the surface where the TF approximation fails.
Hence, a trial function based on the TF approximation prevents the quantitative evaluation of the kinetic energy, which affects the frequency of the quadrupole mode.
To accurately compare $\omega_Q$ with the numerical results, we need to introduce a first-order correction to the kinetic energy of the equilibrium state of the self-gravitating BEC.
This investigation will be conducted in a future study.

\section{CONCLUSIONS}

We study the collective modes of the self-gravitating BECs.
In particular, we focus on the breathing and anisotropic collective modes.
In this study, we show the following three aspects.

First, a self-gravitating BEC can induce a breathing mode by the introduction of an isotropic initial velocity field to its equilibrium state.
Due to the time-dependent density profile of BEC, the gravitational potential near the center of the BEC oscillates in this mode.
The oscillation is characterized by a harmonic oscillation of the radius of the BEC, similar to a conventional BEC trapped by an external potential.  
In the self-gravitating BEC, as the total mass $M$ increases, the amplitude decreases and the period becomes short in proportion to $1/\sqrt{M}$.
These distinctive properties distinctly reflect the density dependence of self-gravity.

Secondly, the axisymmetric initial velocity field yields an anisotropic collective mode.
It is a mixture of the quadrupole mode and the breathing mode.
The quadrupole mode can be extracted by focusing on the oscillation of the aspect ratio of the BEC.  
The amplitude and period in the quadrupole mode decrease as the total mass increases, similar to the radius of BEC in the breathing mode. 
Particularly, the period of the quadrupole mode has the same $M$-dependence, in proportion to $1/\sqrt{M}$, as that of the breathing mode. 
This property of the self-gravitating BEC differs from that of a conventional BEC trapped by an external potential. 
Although the $M$-dependence of the period is different for the self-gravitating BEC and the one in an external isotropic harmonic potential, the ratio of the period of the quadrupole mode to that of the breathing mode takes the same value of $\sqrt{5/2}$ in both cases. 
These properties of the period of the quadrupole mode are also shown analytically in a previous study employing a sum-rule approach \cite{Giovanazzi2001}.

Thirdly, we can also extract only the quadrupole mode from the anisotropic collective mode by setting an appropriate initial velocity.
We extend the variational method, in which the trial function is based on the TF solution, to a spheroidal configuration.
With this approach, we succeed in describing the anisotropic collective mode of the self-gravitating BEC and proposing appropriate initial conditions for the pure quadrupole mode. 
However, this extended variational method does not work for the evaluation of the kinetic energy of the equilibrium state. 
Consequently, the frequency of the quadrupole mode cannot be accurately estimated because it depends critically on the kinetic energy to the equilibrium state.
To quantitatively reproduce the frequency, it is necessary to introduce a first-order correction to the kinetic energy.

In instances where the oscillation exhibits a large amplitude, the density dependency inherent in the self-gravitating BEC probably changes the $M$-dependence of the periods and other characteristics of the collective mode.
In this study, we consider the amplitude to be small.
However, the large deformation of a self-gravitating BEC affects the gravitational energy.
Accounting for this deformation, a collective mode with a large amplitude can cause various nontrivial phenomena triggered by the difference in whether the spheroidal configuration is oblate or prolate.

If the self-gravitating BEC explains a DM halo, we expect that the collective oscillations of the self-gravitating BEC provide evidence that the DM halo is composed of BEC. 
This is a similar situation to early studies on atomic BECs, whose collective mode played an important role in determining whether a system was a BEC.
For instance, the collective modes investigated in the present paper may be caused by the collision and merging of BECs. 
The time scales of these oscillations are shorter than the dynamical time scales of the cluster of galaxies, the breathing modes excite acoustic waves in baryonic matter, and the anisotropic modes, including quadrupole moments, excite gravitational waves.
The frequencies of these oscillations are determined mainly by the self-interaction of the Bose particles, and these frequencies depend on the total mass of the BEC, the boson mass, and the s-wave scattering length.
Hence, these waves provide important information about the Bose particle and the BEC in the universe. 
Observational study of such waves with long-wavelengths is a challenging work.

One of our next interests is the collective modes of rotating self-gravitating BEC. 
In the universe, objects are naturally rotating rather than in static states.
The equilibrium state of a self-gravitating BEC is affected by rotation owing to its density dependence.
For example, the presence of a quantized vortex within a BEC locally pushes the density through rotation, changing the density profile of the equilibrium state.
Since the gravitational potential depends on the density distribution, it differs from that of a BEC without a quantized vortex.
Hence, the self-gravitating BEC with rotation or quantized vortex likely provides distinct phenomena of the collective modes from those in the present work.
Clarifying this property is an interesting next task. 

\section*{Acknowledgment}
We thank M. Kobayashi and Y. Sano for their fruitful discussions of our numerical calculations.
This work was supported by JST and the establishment of university fellowships for the creation of science and technology innovation, Grant Number JPMJFS2138.
M. T. acknowledges the support from JSPS KAKENHI Grant Number JP23K03305.
This study was partially supported by Osaka Central Advanced Mathematical Institute. 
MEXT Joint Usage/Research Center on Mathematics and Theoretical Physics, JPMXP0619217849.

\vspace{0.2cm}
\noindent

\let\doi\relax


\begin{thebibliography}{99}
\bibitem{Rubin1970} V. Rubin and W. Ford, Astrophys. J, \textbf{159}, 379 (1970).
\bibitem{refregier2003} A. Refregier, Annu. Rev. Astron. Astrophys. \textbf{41}, 645 (2003).
\bibitem{Freese2009} K. Freese, EAS, \textbf{36}, 113 (2009).
\bibitem{Binney} J. Binney and S. Tremaine, \it{GALACTIC DYNAMICS}, \rm{second edition}, (Princeton University Press, 2008).
\bibitem{popolo2017} A. D. Popolo and M. L. Delliou, galaxies, \textbf{5}, 1, 17 (2017).
\bibitem{Bullock2017} J. S. Bullock and M. Boylan-Kolchin, Annu. Rev. Astron. Astrophys., \textbf{55}, 343 (2017).
\bibitem{banerjee2022} A. Banerjee, K. K. Boddy, F. -Y. Cyr-Racine, A. L. Erickcek, D. Gilman, B. V. Lehmann, Y. -Y. Mao, P. Mocz, F. Munshi, E. O. Nadler, L. Necib, A. Parikh, A. H. G. Peter, L. Sales, M. Vogelsberger and A. C. Wright, arXiv:2203.07049 (2022).
\bibitem{Hu2000} W. Hu, R. Barkana and A. Gruzinov, Phys. Rev. Lett. \textbf{85}, 1158 (2000).
\bibitem{Hui2021} L. Hui, Annu. Rev. Astron. Astrophys., \textbf{59}, 247 (2021).
\bibitem{spergel2000} D. N. Spergel and P. J. Steinhardt, Phys. Rev. Lett., \textbf{84}, 3760 (2000).
\bibitem{Tulin2018} S. Tulin and H. -B. Yu, Phys. Rep., \textbf{730}, 1 (2018).
\bibitem{magana2012} J. Maga\~{n}a and T. Matos, J. Phys.: Conf. Ser. \textbf{378}, 012012 (2012).
\bibitem{niemeyer2020} J. C. Niemeyer, Progress in Particle and Nuclear Physics \textbf{113} ,103787 (2020).
\bibitem{ferreira2021} E. G. M. Ferreira, Astron. Astrophys. Rev. \textbf{29}, 7 (2021).
\bibitem{PethickSmith} C. J. Pethick and H. Smith, \it{Bose-Einstein Condensation in Dilute Gases}, \rm{second edition}, (Cambridge University Press, 2008).
\bibitem{matos2001} T. Matos and L. A. Ure\~{n}a-L\'{o}pez, Phys. Rev. D \textbf{63}, 063506(2001).
\bibitem{ACE} C. M. Gonz\'{a}lez, J. E. M. Aguilar and L. M. R. Barrera, \textit{Accelerated Cosmic Expansion : Proceedings of the Fourth International Meeting on Gravitation and Cosmology}, \textrm{(Springer, 2013)}.
\bibitem{silverman2002} M. P. Silverman and R. L. Mallett, General relativity and Gravitation \textbf{34}, 633 (2002).
\bibitem{hui2017} L. Hui, J.P. Ostriker, S. Tremaine, E. Witten, Phys. Rev. D \textbf{95},043541 (2017).
\bibitem{marsh2016} D. J. E. Marsh, Phys. Rep., \textbf{643}, 1 (2016).
\bibitem{dalfovo1999} F. Dalfovo, S. Giorgini, L. P. Pitaevskii, and S. Stringari, Rev. Mod. Phys. \textbf{71}, 463 (1999).
\bibitem{fetter2009} A. L. Fetter, Rev. Mod. Phys. \textbf{81}, 647 (2009).
\bibitem{tsatsos2016} M. C. Tsatsos, P. E. S. Tavares, A. Cidrim, A. R. Fritsch, M. A. Caracanhas, F. E. A. dos Santos, C. F. Barenghi and V. S. Bagnato, Phys. Rep. \textbf{622}, 1 (2016).
\bibitem{tsubota2013} M. Tsubota, M. Kobayashi and H. Takeuchi, Phys. Rep. \textbf{522}, 191 (2013).
\bibitem{stringari} L. Pitaevskii and S. Stringari, \textit{Bose-Einstein Condensation}, (Oxford University Press, 2003).
\bibitem{leggett} A. J. Leggett, \textit{Quantum Liquids Bose Condensation and Cooper Pairing in Condensed-Matter Systems}, (Oxford University Press 2006).
\bibitem{Gross1963} E. P. Gross, J. Math. Phys. Rev. \textbf{4}, 195 (1963).
\bibitem{Pitaevskii1961} L. P. Pitaevskii, Sov. Phys. JETP \textbf{13}, 451 (1961).
\bibitem{chavanis2011_1}P. H. Chavanis, Phys. Rev. D \textbf{84}, 043531(2011).
\bibitem{chavanis2011_2}P. H. Chavanis and L. Delfini, Phys. Rev. D \textbf{84}, 043532 (2011).
\bibitem{harvey2015} D. Harvey, R. Massey, T. Kitching, A. Taylor and E. Tittley, Science \textbf{347}, 1462 (2015).
\bibitem{jauzac2016} M. Jauzac, D. Eckert, J. Schwinn, D. Harvey, C. M. Baugh, A. Robertson, S. Bose, R. Massey, M. Owers, H. Ebeling, H. Y. Shan, E. Jullo, J. -P. Kneib. J. Richard, H. Atek, B. Cl\'ement, E. Egami, H. Israel, K. Knowles, M. Limousin, P. Natarajan, M. Rexroth, P. Taylor and C. Tchernin, Mon. Not. R. Astron. Soc. \textbf{463}, 3876 (2016).
\bibitem{bohmer2007} C. G. B\"{o}hmer and T. Harko, JCAP \textbf{06}, 025(2007).
\bibitem{harko2011} T. Harko, J. Cosmol. Astropart. Phys. \textbf{05}, 022 (2011).
\bibitem{harko2015} T. Harko and F. S. N. Lobo, Phys. Rev. D \textbf{92}, 043011 (2015).
\bibitem{zhang2018} X. Zhang, M. H. Chan, T. Harko, S. -D. Liang and C. S. Leung, Eur. Phys. J. C \textbf{78}, 346 (2018).
\bibitem{kain2010} B. Kain and H. Y. Ling, Phys. Rev. D \textbf{82}, 064042 (2010).
\bibitem{daller2012} T. Rindler-Daller and P. R. Shapiro, MNRAS \textbf{422}, 135 (2012).
\bibitem{harko2014} T. Harko, Phys. Rev. D \textbf{89}, 084040 (2014).
\bibitem{chavanis2016} P.-H. Chavanis, Phys. Rev. D \textbf{94}, 083007 (2016).
\bibitem{chavanis2018} P.-H. Chavanis, Phys. Rev. D \textbf{98}, 023009 (2018).
\bibitem{guzman2013} F. S. Guzm\'{a}n, F. D. Lora-Clavijo, J. J. Gonz\'{a}lez-Avil\'{e}s and F. J. Rivera-Paleo, J. Cosmol. Astropart. Phys. \textbf{09}, 034 (2013).
\bibitem{guzman2006} F. S. Guzm\'{a}n and L. A. Ure\~{n}a-L\'{o}pez, Astrophys. J. \textbf{645}, 814 (2006).
\bibitem{guzman2014} F. S. Guzm\'{a}n, F. D. Lora-Clavijo, J. J. Gonz\'{a}lez-Avil\'{e}s and F. J. Rivera-Paleo, Phys. Rev. D \textbf{89}, 063507 (2014).
\bibitem{guzman2015} F. S. Guzm\'{a}n, F. D. Lora-Clavijo, Gen. Relativ. Gravit. \textbf{47}, 21 (2015).
\bibitem{gonzalez2011} J. A. Gonz\'alez and F. S. Guzm\'an, Phys. Rev. D \textbf{83}, 103513 (2011).
\bibitem{nikolaieva2021} Y. O. Nikolaieva, A. O. Olashyn, Y. I. Kuriatnikov, S. I. Vilchinskii and A. I. Yakimenko, Low Temperature Physics \textbf{47}, 684 (2021).
\bibitem{dmitriev2021} A. S. Dmitriev, D. G. Levkov, A. G. Panin, E. K. Pushnaya and I. I. Tkachev, Phys. Rev. D \textbf{104}, 023504 (2021).
\bibitem{nikolaieva2022} Y. O. Nikolaieva, Y. M. Bidasyuk, K. Korshynska, E. V. Gorbar, J. Jia and A. I. Yakimenko, Phys. Rev. D \textbf{108}, 023503 (2023).
\bibitem{harko2019} T. Harko, Eur. Phys. J. C \textbf{79}, 787 (2019).
\hypertarget{Thijssen}{\bibitem{Thijssen}} J. M. Thijssen, \textit{Computational Physics}, (Cambridge University Press, 2007).
\bibitem{nesterov1983}Y. Nesterov, Soviet Math. Dokl. 27, 372 (1983).
\bibitem{su2016} W. Su, S. Boyd and E. J. Cand\`{e}s, J. Mach. Learn. Res. \textbf{17}, 153 (2016).
\bibitem{donoghue2015} B. O'Donoghue and E. J. Cand\`{e}s, Found. Comput. Math. \textbf{15}, 715 (2015).
\bibitem{castin1996}Y. Castin and R. Dum, Phys. Rev. Lett. \textbf{77}, 5315 (1996).
\bibitem{hayashi2021}K. Hayashi, M. Ibe, S. Kobayashi, Y. Nakayama and S. Shirai, Phys. Rev. D \textbf{103}, 023017 (2021).
\bibitem{Jin1996}D. S. Jin, J. R. Ensher, M. R. Matthews, C. E. Wieman and E. A. Cornell, Phys. Rev. Lett. \textbf{77}, 420 (1996).
\bibitem{Mewes1996}M. -O. Mewes, M. R. Andrews, N. J. van Druten, D. M. Kurn, D. S. Durfee, C. G. Townsend and W. Ketterle, Phys. Rev. Lett. \textbf{77}, 988 (1996).
\bibitem{Deppner2021}C. Deppner, W. Herr, M. Cornelius, P. Stromberger, T. Sternke, C. Grzeschik, A. Grote, J. Rudolph, S. Herrmann, M. Krutzik, A. Wenzlawski, R. Corgier, E. Charron, D. Gu\'{e}ry-Odelin, N. Gaaloul, C. L\"{a}mmerzahl, A. Peters, P. Windpassinger and E. M. Rasel, Phys. Rev. Lett. \textbf{127}, 100401 (2021).
\bibitem{Giovanazzi2001} S. Giovanazzi, G. Kurizki, I. E. Mazets, and S. Stringari, Europhys. Lett., \textbf{56}, 1 (2001).
\bibitem{lai1993} D. Lai, F. A. Rasio and S. L. Shapiro, ApJS, 88, 205 (1993).
\end{thebibliography}
\end{document}